\documentclass[aps,prl,twocolumn,showpacs,superscriptaddress,floatfix,10pt]{revtex4-1}

\usepackage{amssymb}  
\usepackage{color,soul} 
\usepackage{graphicx}  
\usepackage[tight]{subfigure}  
\usepackage{mathrsfs}
\usepackage[pdftex,pdfusetitle,bookmarks=true,colorlinks=true,citecolor=blue,urlcolor=blue,linkcolor=magenta]{hyperref}
\usepackage{hypcap}
\usepackage{braket}
\usepackage{xfrac}
\usepackage{tikz}
\usepackage{url}
\usepackage{setspace}
\usepackage{enumerate}
\usepackage{bm}
\usepackage{epstopdf}
\usepackage{amsmath}
\usepackage{amsfonts}
\usepackage{mhchem} 
\usepackage[normalem]{ulem}
\usepackage{tikz-feynman}

\begin{document}

\title{Collective spinon spin wave in a magnetized U(1) spin liquid}
\author{Leon Balents}
\affiliation{Kavli Institute for Theoretical Physics, University of California, Santa Barbara, CA 93106, USA}
\author{ Oleg A. Starykh}
\affiliation{Department of Physics and Astronomy, University of Utah, Salt Lake City, Utah 84112, USA}

\begin{abstract} \label{abstract}
  We study the transverse dynamical
  spin susceptibility 
  of the two dimensional U(1) spinon Fermi surface spin liquid in a
  small applied Zeeman field. We show that both short-range
  interactions, present in a generic Fermi liquid, as well as gauge
  fluctuations, characteristic of the U(1) spin liquid, qualitatively
  change the result based on the frequently assumed non-interacting
  spinon approximation.  The short-range part of the interaction leads
  to a new collective mode: a ``spinon spin wave'' which splits off
  from the two-spinon continuum at small momentum and disperses
  downward.  Gauge fluctuations renormalize the susceptibility,
  providing non-zero power law weight in the region outside the spinon
  continuum
  and giving the spin wave a finite lifetime, which scales as momentum squared.
  We also study the effect of Dzyaloshinskii-Moriya anisotropy on
  the zero momentum susceptibility, which is measured in electron spin
  resonance (ESR), and obtain a resonance linewidth linear in temperature
  and varying as $B^{2/3}$ with magnetic field $B$ at low
  temperature.  Our results form the basis for a theory of inelastic
  neutrons scattering, ESR, and resonant inelastic
  x-ray scattering (RIXS) studies of this quantum spin liquid state.
\end{abstract}
\date{\today}

\maketitle

The search for the enigmatic spin liquid state has switched into high gear in recent years. Dramatic theoretical (Kitaev model \cite{Kitaev2006,Savary2017} and spin liquid in triangular lattice antiferromagnet \cite{Zhu2015}) and experimental (YbMgGaO$_4$ \cite{Paddison2017,Shen2018} and $\alpha$-RuCl$_3$ \cite{Banerjee2016}) developments leave no doubt of the eventual success of this enterprise.  To push this to the next stage, it is incumbent upon the community to identify specific experimental signatures that evince the unique aspects of these states.  In this paper, we focus on the two dimensional $U(1)$ quantum spin liquid (QSL) with a spinon Fermi surface. This is a priori the most exotic two dimensional QSL state, and yet one which has repeatedly been advocated for in both theory \cite{Ioffe1989,Motrunich2005,Lee2005,Motrunich2006,Nave2007} and experiment \cite{Yamashita2008,Shen2016,Fak2017,Shen2018}.  Specifically, we study the  dynamical susceptibility of the ${\bf q}$-component of the spin operator $S^a_{\bf q} ~(a=x,y,z)$
\begin{equation}
  \label{eq:2}
  X_\pm(\bm{q},\omega) = - i \int_0^\infty dt \langle [S^+_{\bf q}(t), S^-_{-{\bf q}}(0)]\rangle e^{i \omega t},
\end{equation}
which is an extremely information-rich quantity, and is accessible
through inelastic neutron scattering \cite{Banerjee2017}, ESR
\cite{Smirnov2015,Ponomaryov2017}, and RIXS \cite{Halasz2016}.  
The fractionalization of triplet excitations into pairs of spinons is
a fundamental aspect of a QSL, and is expected to
manifest in  $X_\pm$ as two-particle continuum spectral weight \cite{Norman2016,Savary2017,Ng2017}, a
surprising feature which appears more characteristic of a weakly
correlated metal than a strongly correlated Mott insulator.

\begin{figure}[h]
       \includegraphics[width=7.cm]{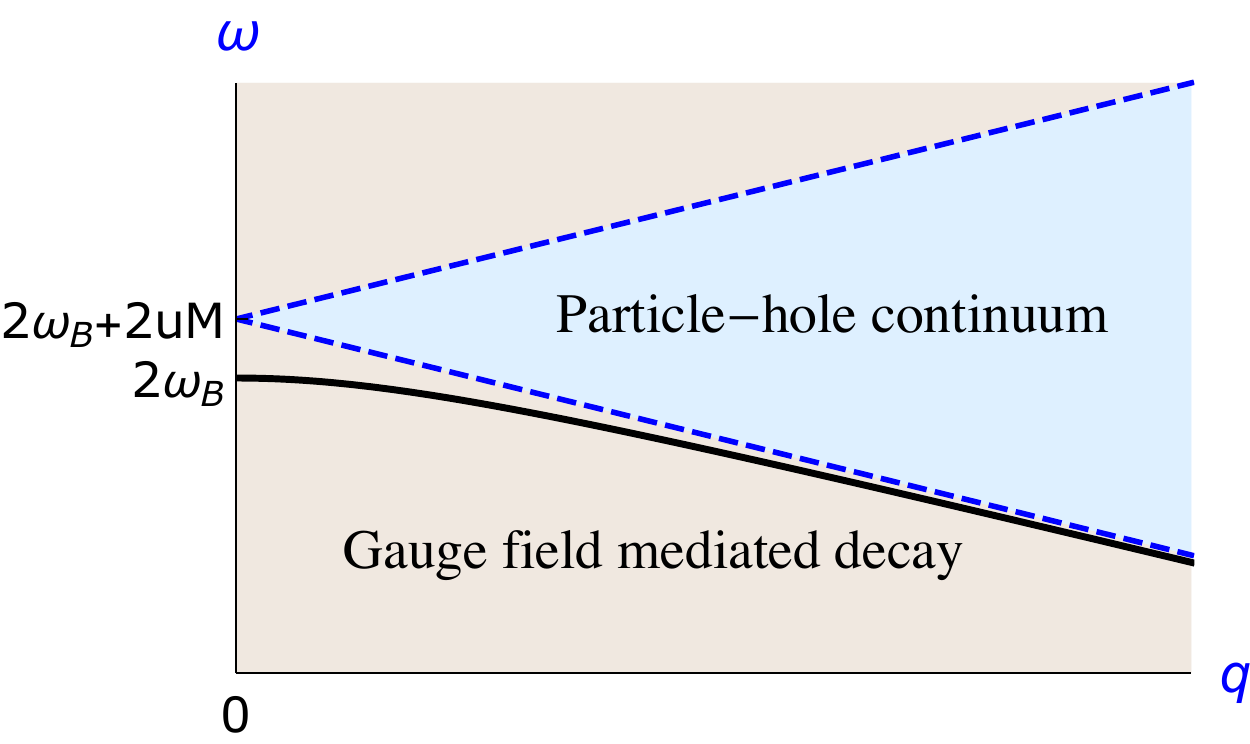}
    \caption{Magnetic excitation spectrum of an interacting U(1) spin liquid with spinon Fermi surface. Wedge-shaped (blue) region, bounded by dashed lines,
     denotes spinon particle-hole continuum inside which $\Im\chi^0_\pm(\bm{q},\omega) \neq 0$. Collective spinon spin wave, which is promoted by short-ranged repulsive interaction $u$, is shown by the bold black line. The linewidth of this transverse spin wave is determined by the gauge fluctuations, which produce finite 
   $\Im\chi^1_\pm(\bm{q},\omega) \neq 0$ outside the non-interacting spinon continuum, see \eqref{eq:100}. }
     \label{fig:continuum}
\end{figure}
In a mean-field treatment in which the spinons are approximated as
non-interacting fermions, this continuum has a characteristic shape at
small frequency and wavevector in the presence of an applied Zeeman
magnetic field, as discussed in \cite{Li2017}.  In particular, there
is non-zero spectral weight in a wedge-shaped region which terminates
at a single point along the energy axis at zero momentum.  Our
analysis reveals the full structure in this regime beyond the mean
field approximation.  Notably, we find that interactions between
spinons {\em qualitatively} modify the result from the mean-field
form, introducing a new collective mode -- a {\em ``spinon spin wave''} --
and modifying the spectral weight significantly.

We recapitulate the derivation of the theory of the spinon Fermi surface phase \cite{Nagaosa1999,Lee2006}.  One introduces Abrikosov fermions by rewriting the spin operator  $\bm{S}_i = \frac{1}{2} c_{i\alpha}^\dagger \bm{\sigma}_{\alpha\beta} c_{i\beta}^{\vphantom\dagger}$, where $c_{i\alpha}^{\vphantom\dagger},c_{i\alpha}^\dagger$ are canonical fermionic spinors on site $i$ with spin-1/2 index $\alpha$ (repeated spin indices are summed).  This is a faithful representation provided the constraint $c_{i\alpha}^\dagger c_{i\alpha}^{\vphantom\dagger}=1$ is imposed -- this constraint induces a gauge symmetry.    In a path integral representation, the constraint is enforced by a Lagrange multiplier $A_{i0}$, which takes the role of the time-component of a gauge field, i.e. scalar potential.  Microscopic exchange interactions, which are quadratic in spins, and are therefore quartic in fermions, are decoupled to introduce new link fields whose phases act as the spatial components of the corresponding gauge fields ${\bf A}$, i.e. the vector potential.  

To describe the universal low energy physics, it is appropriate to
consider ``coarse-grained'' fields
$\psi_\alpha^{\vphantom\dagger},\psi_\alpha^\dagger$ descending from
the microscopic ones, and include the symmetry-allowed Maxwell terms
for the U(1) gauge field.  Furthermore, due to the finite density of
states at the spinon Fermi surface, the longitudinal scalar potential
is screened and the time component $A_0$ can then be integrated out to
mediate a short-range repulsive interaction $u$ between like charges.  Therefore we consider the Euclidean action $S = S_\psi + S_A + S_u$, where \cite{Kim1994,Nagaosa1999,Lee2006}
\begin{eqnarray}
  \label{eq:1}
  &&S_\psi = \int \!d^3 x\,  \psi^{\dagger} \Big(\partial_\tau - \mu  - \frac{1}{2m}({\bm \nabla}_{\bf r} - i {\bf A})^2- \omega_B \sigma^3\Big)\psi,\nonumber\\
  &&S_A = \int \!\frac{d^3q}{(2\pi)^3} \, \frac{1}{2}   \left( \gamma|\omega_n|/q + \chi q^2\right) |A(q)|^2,\nonumber\\ 
  &&S_u = \int\! d^3x\, u \, \psi_\uparrow^{\dagger} \psi^{\vphantom\dagger}_\uparrow \psi_\downarrow^{\dagger} \psi^{\vphantom\dagger}_\downarrow.
\end{eqnarray}
Here $x=(\tau,\bm{x})$ is the space-time coordinate,
$q=(\omega_n,\bm{q})$ is the three-momentum, $\psi_\alpha$ is a
two-component spinor, with spin indices $\alpha,\beta =
\uparrow,\downarrow$ that are suppressed when possible, $\omega_B$ describes static magnetic field ${\bf B} = B\hat{z}$ and includes the g-factor as well as the Bohr magneton.   The gauge dynamics is derived in the 
Coulomb gauge ${\bm \nabla}\cdot {\bm A} = 0$ with $\bm{A}(q) =
i\bm{\hat{z}}\times \bm{\hat{q}}A(q)$. The gauge action $S_A$ is
generated by spinons and $\gamma = 2 \bar{n}/k_F$ and $\chi = 1/(24\pi
m)$ represent Landau damping and diamagnetic susceptibility of
non-interacting spinon gas, correspondingly ($m$ is the spinon mass,
$\bar{n}$ is the spinon density and $k_F$ is the Fermi momentum of
non-magnetized system). 

We proceed with the assumption of SU(2) symmetry, a good first
approximation for many spin liquid materials, and address the effect of
its violations in the latter part of the paper.  Previous
investigations focused on the transverse vector potential ${\bf A}$,
which is not screened but Landau damped, and hence induces exotic
non-Fermi-liquid physics.  For example, one finds a self-energy
varying with frequency as $\omega^{2/3}$, and a singular contribution
to the specific heat $c_v \sim T^{2/3}$ \cite{Nagaosa1999,Lee2006}.  However, notably, the
transverse gauge field has negligible effects on the hydrodynamic
long-wavelength collective response \cite{Kim1994}.  Here, we instead
focus on the short-range repulsion $u$, which produces an exchange field
that dramatically alters the behavior in the presence of an external
Zeeman magnetic field giving rise to finite magnetization.  Gauge fluctuations play a
subsidiary role which we also include.

An important constraint follows purely from symmetry.  Provided the Hamiltonian in zero magnetic field has SU(2) symmetry, a Zeeman magnetic field leads to a fully determined structure factor at zero momentum.  Specifically, the Larmor/Kohn theorem \cite{Oshikawa2002} dictates that the only response at 
${\bf q}=0$, $X_\pm'' = -2M \delta(\omega - 2 \omega_B)$, where $M=(\bar{n}_\uparrow - \bar{n}_\downarrow)/2$ is the magnetization and $\omega_B$ is the spinon Zeeman energy.  For free fermions, the delta function is precisely at the corner of the {\em spinon} particle-hole continuum (also known as the two-spinon continuum).  
However, the contact exchange interaction shifts up the particle-hole continuum, at small momentum ${\bf q}$, away from the Zeeman energy $2 \omega_B$ to $2\omega_B + 2 u M$.   This is seen by the trivial Hartree self-energy
\begin{equation}
  \label{eq:52}
  \Sigma_\sigma =
  \begin{tikzpicture}[baseline={([yshift=-3.0ex]current bounding box.center)}]
    \tikzset{
      every node/.append style={font=\small},
      every edge/.append style={thick},
      arrow/.style={thick, shorten >=5pt,shorten <=5pt,->},
      photon/.style={decorate, draw=blue,decoration={snake}},
      electron/.style={thick,postaction={decorate},
        decoration={markings,mark=at position .55 with {\arrow[]{>}}}},
      fermion/.style={thin,postaction={decorate},
        decoration={markings,mark=at position .7 with {\arrow[]{>}}}},
      gluon/.style={decorate, draw=magenta,
        decoration={zigzag,amplitude=2pt, segment length=5pt}},
    }
    \draw [fermion] (2.7,0) to (2.9,0);
    \draw [fermion] (2.9,0) to (3.1,0);
    \fill (2.9,0) circle (1pt) coordinate(ar);
    \fill (2.9,0.5) circle (1pt) coordinate(ar2);
    \fill (2.2,1.0) node[right] {$-\sigma$};
    \draw [gluon] (ar)--(ar2);
    \begin{scope}[thick,decoration={
        markings, mark=at position 0.28 with {\arrow{>}}}
      ]
      \draw[postaction={decorate}] (2.9,0.7) circle (0.2cm);
    \end{scope}
  \end{tikzpicture}
  = u \bar{n}_{-\sigma} = -  u M\sigma + u \bar{n} /2,
\end{equation}
where we use a zig-zag line to diagrammatically represent the local $u$ interaction, $\sigma=\uparrow=1$ and $\sigma=\downarrow=-1$, and
$\bar{n}_\sigma$ is the expectation value of spin-$\sigma$ spinon
density in the presence of magnetic field. Consequently,
for the Larmor theorem to be obeyed, there {\em must be a collective
  transverse spin mode} at small momenta.

This collective spin mode is most conveniently described by the Random Phase Approximation (RPA), which corresponds to a standard resummation of particle-hole ladder diagrams \cite{aronov1977}.  
For the particular case of a momentum-independent contact interaction, one has
\begin{align}
  \label{eq:3}
  X_\pm(\bm{q},i\omega_n) & =
                               \begin{tikzpicture}[baseline={([yshift=-0.5ex]current bounding box.center)}]
     \tikzset{
      every node/.append style={font=\small},
      every edge/.append style={thick},
      arrow/.style={thick, shorten >=5pt,shorten <=5pt,->},
      photon/.style={decorate, draw=blue,decoration={snake}},
      electron/.style={thick,postaction={decorate},
        decoration={markings,mark=at position .55 with {\arrow[]{>}}}},
      fermion/.style={thin,postaction={decorate},
        decoration={markings,mark=at position .7 with {\arrow[]{>}}}},
      gluon/.style={decorate, draw=magenta,
        decoration={zigzag,amplitude=1pt, segment length=3pt}},}
    \coordinate (in) at (-.1,0);
    \coordinate (out) at (1.5,0);
    \draw (2.6,1) circle (1pt) coordinate (lt3);
    \draw (3.6,1) circle (1pt) coordinate (rt3);   
    \draw (lt3) edge[electron,out=45,in=135] node[above] {$~~\uparrow$} (rt3);
    \draw (rt3) edge[electron,out=-135,in=-45] node[below] {$~~\downarrow$} (lt3);
  \end{tikzpicture}+
\begin{tikzpicture}[baseline={([yshift=-0.5ex]current bounding box.center)}]
     \tikzset{
      every node/.append style={font=\small},
      every edge/.append style={thick},
      arrow/.style={thick, shorten >=5pt,shorten <=5pt,->},
      photon/.style={decorate, draw=blue,decoration={snake}},
      electron/.style={thick,postaction={decorate},
        decoration={markings,mark=at position .55 with {\arrow[]{>}}}},
      fermion/.style={thin,postaction={decorate},
        decoration={markings,mark=at position .7 with {\arrow[]{>}}}},
      gluon/.style={decorate, draw=magenta,
        decoration={zigzag,amplitude=1pt, segment length=3pt}},}
    \coordinate (in) at (-.1,0);
    \coordinate (out) at (1.5,0);
    \draw (2.6,1) circle (1pt) coordinate (lt3);
    \draw (3.6,1) circle (1pt) coordinate (rt3);   
    \fill (2.9,0.83) circle (1pt) coordinate (oob2);
    \fill (2.9,1.17) circle (1pt) coordinate (at2);
    \draw (lt3) edge[electron,out=45,in=135] node[above] {$~~\uparrow$} (rt3);
    \draw (rt3) edge[electron,out=-135,in=-45] node[below] {$~~\downarrow$} (lt3);
    \draw [gluon] (at2) to (oob2);
  \end{tikzpicture}+
                               \begin{tikzpicture}[baseline={([yshift=-0.5ex]current bounding box.center)}]
     \tikzset{
      every node/.append style={font=\small},
      every edge/.append style={thick},
      arrow/.style={thick, shorten >=5pt,shorten <=5pt,->},
      photon/.style={decorate, draw=blue,decoration={snake}},
      electron/.style={thick,postaction={decorate},
        decoration={markings,mark=at position .55 with {\arrow[]{>}}}},
      fermion/.style={thin,postaction={decorate},
        decoration={markings,mark=at position .7 with {\arrow[]{>}}}},
      gluon/.style={decorate, draw=magenta,
        decoration={zigzag,amplitude=1pt, segment length=3pt}},}
    \coordinate (in) at (-.1,0);
    \coordinate (out) at (1.5,0);
    \draw (2.6,1) circle (1pt) coordinate (lt3);
    \draw (3.6,1) circle (1pt) coordinate (rt3);   
    \fill (2.9,0.83) circle (1pt) coordinate (oob2);
    \fill (2.9,1.17) circle (1pt) coordinate (at2);
    \fill (3.3,1.17) circle (1pt) coordinate (oot);
    \fill (3.3, 0.83) circle (1pt) coordinate (ab);    
    \draw (lt3) edge[electron,out=45,in=135] node[above] {$~~\uparrow$} (rt3);
    \draw (rt3) edge[electron,out=-135,in=-45] node[below] {$~~\downarrow$} (lt3);
    \draw [gluon] (at2) to (oob2);
    \draw [gluon] (oot) to (ab);
  \end{tikzpicture}+\cdots \nonumber \\
  & =  \frac{\chi_\pm(\bm{q},i\omega_n)}{1 + u \chi_\pm(\bm{q},i\omega_n)},
\end{align}
where the fermion lines correspond to the spinon Green's functions
{\em including the Hartree shift} \eqref{eq:52}, and in this approximation
$\chi_\pm(\bm{q},i\omega_n)=\chi^0_\pm(\bm{q},i\omega_n) $ is the bare
susceptibility bubble, calculated using these functions.  We will
however use the second line in Eq.~\eqref{eq:3} to later define the
RPA approximation even when gauge field corrections (but not the local
interaction $u$) are included in $\chi_\pm$.  For the moment, we
simply evaluate the bare susceptibility,
\begin{eqnarray}
  \label{eq:4}
  \chi^0_\pm(\bm{q},i\omega_n) &=& \frac{1}{\beta V} \sum_{k_n,\bm{k}} \frac{1}{ik_n -\epsilon_{\bm{k}} +\omega_B - u \bar{n}_\downarrow} 
 \nonumber\\&&\times \frac{1}{ik_n +i\omega_n -\epsilon_{\bm{k}+\bm{q}} -\omega_B - u \bar{n}_\uparrow} .
\end{eqnarray}
Here $\omega_n,k_n$ are bosonic and fermionic Matsubara frequencies,
respectively.  A simple calculation, followed by analytical continuation $i\omega_n \to \omega + i0$, gives
\begin{eqnarray}
  \label{eq:5}
\Re\chi^0_\pm(\bm{q},\omega) = \frac{2M {\rm sign}(\omega-2\omega_B - 2uM)}{\sqrt{(\omega-2\omega_B - 2uM)^2 -v_F^2 q^2}},\nonumber\\
\Im\chi^0_\pm(\bm{q},\omega) = \frac{-2M}{\sqrt{v_F^2q^2-(\omega-2\omega_B - 2uM)^2}},
\end{eqnarray}
where square-roots are defined when their arguments are positive.
The real/imaginary spin susceptibility describes domains outside/inside two-spinon continuum in the $(q,\omega)$ plane, Fig.\ref{fig:continuum}, 
correspondingly. At $\bm{q}=0$
\begin{equation}
  \label{eq:6}
  \chi^0_\pm(\bm{q}=0,\omega) = \frac{2M}{\omega - 2\omega_B - 2u M + i0},
\end{equation}
and therefore $\Im \chi^0_\pm(\bm{q}=0,\omega) \sim \delta(\omega - 2\omega_B - 2u M)$:
the position of the two-spinon continuum is renormalized by the interaction shift.  However, inserting \eqref{eq:6} in the RPA formula \eqref{eq:3} one finds
that the RPA successfully recovers Larmor theorem at zero momentum for the interacting $SU(2)$-invariant system,
\begin{equation}
  \label{eq:7}
  X_\pm(\bm{q}=0,\omega) = \frac{2M}{\omega - 2\omega_B + i0}.
\end{equation}
 Therefore the contribution at $\bm{q}=0$ is {\em solely} from the collective mode, with no spectral weight from the continuum at $2\omega_B + 2u M$.  
 Dispersion of the collective spin mode is obtained with the help of \eqref{eq:5} and $\Im X_\pm = \Im \chi^0_\pm/[(1+u\Re\chi^0_\pm)^2 + (u \Im\chi^0_\pm)^2]$, 
 \begin{equation}
 \label{eq:8}
 \omega_{\rm swave}({\bf q})=2\omega_B + 2u M - \sqrt{4u^2 M^2 + v_F^2 q^2}.
 \end{equation}
 For small $q \ll u M/v_F$ the collective mode is dispersing downward
 quadratically $\omega \approx 2\omega_B - (v_F q)^2/(4 u M)$, while
 in the opposite limit $q \gg u M/v_F$ it approaches the low boundary
 of the spinon continuum,
 $\omega \approx 2\omega_B + 2u M - v_F q$. Retaining quadratic in $q$
 terms in \eqref{eq:4} will lead to the termination of the collective
 mode at some $q_{\rm max}$ at which the spin wave enters the
 two-spinon continuum.

This physics is not unique to spin liquids but applies to paramagnetic metals.  Historically, this spin wave mode was predicted by Silin in 1958 for non-ferromagnetic metals  within Landau Fermi liquid theory \cite{Silin1958,Silin1959,Platzman1967,Leggett1970}, and observed via conduction electron spin resonance (CESR) in 1967 \cite{Schultz1967}. At the time, this observation was 
considered to be one of the first proofs of the validity of the Landau theory of Fermi-liquids \cite{Platzman1973}.
Unlike the more well-known zero sound \cite{abel1966}, an external magnetic field is required in order to shift the particle-hole continuum up along the energy axis to allow for the undamped collective spin wave to appear outside the particle-hole continuum, in the triangle-shaped window below it. Second order in the interaction $u$ corrections (beyond the ladder series) do cause damping of this spin mode \cite{Musaelian1994,Golosov1995,maslov2018}.

However, in the $U(1)$ spin liquid, there is an additional branch of low energy excitations due to the gauge field ${\bm A}$, dispersing as $\omega \sim q^3$.  The very flat dispersion of the gauge excitations suggests it may act as a momentum sink, so that, for example, an excitation consisting of a particle-hole pair plus a gauge quantum may exist in the ``forbidden'' region where the bare particle-hole continuum vanishes and the collective spin mode lives.  It is therefore critical to understand the effect of the gauge interactions upon the dynamical susceptibility.  
To this end, we consider the dressing of the particle-hole bubble
$\chi^0$ by gauge propagators.  Guided by the above thinking, we
expect that it is sufficient to consider all diagrams with a single
gauge propagator (denoted by wavy line).   
\begin{align}
\label{eq:9}
  \chi^1_\pm(\bm{q},&i\omega_n)  =
  \begin{tikzpicture}[baseline={([yshift=-0.5ex]current bounding box.center)}]
    \tikzset{
      every node/.append style={font=\small},
      every edge/.append style={thick},
      arrow/.style={thick, shorten >=5pt,shorten <=5pt,->},
      photon/.style={decorate,
        draw=blue,decoration={snake,amplitude=1,segment length=3pt}},
      electron/.style={thick,postaction={decorate},
        decoration={markings,mark=at position .55 with {\arrow[]{>}}}},
      fermion/.style={thin,postaction={decorate},
        decoration={markings,mark=at position .7 with {\arrow[]{>}}}},
      gluon/.style={decorate, draw=magenta,
        decoration={zigzag,amplitude=1pt, segment length=3pt}},}
    \draw (0,1) circle (1pt) coordinate (lt);
    \draw (1,1) circle (1pt) coordinate (rt);
    \fill (0.3,1.17) circle (1pt) coordinate (ot);
    \fill (0.7,1.17) circle (1pt) coordinate (ot2);
    \draw (lt) edge[electron,out=45,in=135] node[above] {$~~\uparrow$} (rt);
    \draw (rt) edge[electron,out=-135,in=-45] node[below] {$~~\downarrow$} (lt);
    \draw [photon,out=-45,in=-135] (ot) to (ot2);    
  \end{tikzpicture}+
     \begin{tikzpicture}[baseline={([yshift=-0.5ex]current bounding box.center)}]
    \tikzset{
      every node/.append style={font=\small},
      every edge/.append style={thick},
      arrow/.style={thick, shorten >=5pt,shorten <=5pt,->},
      photon/.style={decorate,
        draw=blue,decoration={snake,amplitude=1,segment length=3pt}},
      electron/.style={thick,postaction={decorate},
        decoration={markings,mark=at position .55 with {\arrow[]{>}}}},
      fermion/.style={thin,postaction={decorate},
        decoration={markings,mark=at position .7 with {\arrow[]{>}}}},
      gluon/.style={decorate, draw=magenta,
        decoration={zigzag,amplitude=1pt, segment length=3pt}},}
    \draw (0,1) circle (1pt) coordinate (lt);
    \draw (1,1) circle (1pt) coordinate (rt);
    \fill (0.3,0.83) circle (1pt) coordinate (ot);
    \fill (0.7,0.83) circle (1pt) coordinate (ot2);
    \draw (lt) edge[electron,out=45,in=135] node[above] {$~~\uparrow$} (rt);
    \draw (rt) edge[electron,out=-135,in=-45] node[below] {$~~\downarrow$} (lt);
    \draw [photon,out=45,in=135] (ot) to (ot2);    
  \end{tikzpicture}
                                 +
                                 \begin{tikzpicture}[baseline={([yshift=-0.5ex]current bounding box.center)}]
    \tikzset{
      every node/.append style={font=\small},
      every edge/.append style={thick},
      arrow/.style={thick, shorten >=5pt,shorten <=5pt,->},
      photon/.style={decorate,
        draw=blue,decoration={snake,amplitude=1,segment length=3pt}},
      electron/.style={thick,postaction={decorate},
        decoration={markings,mark=at position .55 with {\arrow[]{>}}}},
      fermion/.style={thin,postaction={decorate},
        decoration={markings,mark=at position .7 with {\arrow[]{>}}}},
      gluon/.style={decorate, draw=magenta,
        decoration={zigzag,amplitude=1pt, segment length=3pt}},}
    \draw (0,1) circle (1pt) coordinate (lt);
    \draw (1,1) circle (1pt) coordinate (rt);
    \fill (0.3,1.17) circle (1pt) coordinate (ot);
    \fill (0.7,.83) circle (1pt) coordinate (ot2);
    \draw (lt) edge[electron,out=45,in=135] node[above] {$~~\uparrow$} (rt);
    \draw (rt) edge[electron,out=-135,in=-45] node[below] {$~~\downarrow$} (lt);
    \draw [photon] (ot) to (ot2);    
  \end{tikzpicture} 
\end{align}
Calculations described in \cite{SM} lead to 
\begin{equation}
  \label{eq:100}
  {\Im}  \, \chi^1(\bm{q},\omega) = -\frac{\sqrt{3} \gamma^{1/3} k_F}{56\pi^2 \chi^{4/3}} \frac{q^2 \omega^{7/3}}{(\omega - 2\tilde{\omega}_B)^4}.
\end{equation}                            
That is, the dressed susceptibility has a non-zero imaginary part in the previously kinematically forbidden region outside
the spinon particle-hole continuum, see Fig.\ref{fig:continuum}.  This is a new continuum weight.
However, the weight in this new continuum contribution vanishes
quadratically in momentum as $\bm{q}=0$ is approached.  This is an
important check on the calculations, since the Larmor theorem still
applies to the full theory \eqref{eq:1} with the gauge field, which implies that
precisely at zero momentum, there can be no new contributions.   
Similar to Kim {\em et al.} \cite{Kim1994}, who considered diagrams for the density
correlations and optical conductivity, this result relies on important
cancellations between self energy (first two diagrams) and vertex
corrections (last diagram), which are
needed to obtain this proper behavior of $\chi^1$. 

Within the RPA approximation of Eq.~\eqref{eq:3}, but now with $\chi_\pm = \chi^0_\pm + \chi^1_\pm$,
we see that the $q^2$ dependence of ${\Im} \chi^1_\pm$ is sufficient to
ensure that the width (in energy) of the collective spin mode becomes
narrow compared to its frequency at small momentum: this is the
standard criteria for sharpness and observability of a collective
excitation. The real part of $\chi^1$, derived in Eq.~(S53),
modifies the dispersion of the collective mode too, but preserves its
downward ${\bf q}^2$ character within the $1/N$ approximation\cite{SM}. The final result for the dynamical
susceptibility is summarized in Fig.~\ref{fig:continuum}.  Away from
the zero momentum axis $\Im X_\pm(\bm{q},\omega)$ is always non-zero,
and is the sum of several distinct contributions.  Inside this spinon-gauge
continuum, the spinon spin wave appears as a resonance which is
asymptotically sharp at small momentum.  We note that, while our calculations are done in two dimensions, a spinon spin wave with the same qualitative
features is also present in the three dimensional U(1) QSL.

For observation of the spinon spin wave via inelastic neutron or RIXS
experiment, the mode should be present over a range of momenta which
is not too narrow.  Because the extent of the `decay-free" triangular
shaped region in Fig.\ref{fig:continuum} is determined by
$2\omega_B/v_F \sim \sqrt{m \omega_B (\omega_B/E_F)}$, this requires
that the Zeeman energy should be a substantial fraction of the
exchange integral (effective Fermi energy).  This makes spin liquid materials with
$J$ of order 10 K ideal candidates for observation, in constrast with
usual metals for which $\omega_B/E_F$ is vanishingly small.

The above results apply to the case in which SU(2) spin rotation
symmetry is broken only by the applied Zeeman field.  Breaking of the
SU(2) invariance by anisotropies invalidates the Larmor theorem and
causes a shift and more importantly a broadening of the spin
collective mode even at zero momentum \cite{Maiti2015}.  This is of particular
importance for electron spin resonance, which has high energy
resolution but measures directly at zero momentum only \cite{Oshikawa2002}.  The way in
which the resonance is broadened depends in detail on the nature of
the anisotropy, the orientation of the applied magnetic field, etc.,
so it is not possible to give a single general result.  Instead, we
provide one example of this physics and consider the
influence of a Dzyaloshinskii-Moriya (DM) interaction, which is typically the dominant form of anisotropy for weakly spin-orbit coupled systems, 
provided it is symmetry allowed by the lattice.  See for example \cite{Smith2003,Winter2017}.

Guided by the arguments of symmetry and simplicity we next suppose
that when projected to the spin liquid ground state manifold of the
two-dimensional spin model, the DM interaction appears as a familiar
spin-orbit interaction of the Rashba kind.  A
momentum-dependent spin splitting, of which this is the simplest
example, is expected to appear in a model without spatial inversion
symmetry because in the spinon Fermi surface state the spinons
transform under lattice point group symmetries in the same way as
usual electrons \cite{Iaconis2018}.  
Then the term in the Hamiltonian breaking SU(2) spin invariance, $H'$, reads 
\begin{equation}
H' = \int d^2 \bm{x} ~\alpha_{\rm so} \psi^\dagger ((\hat{p}_x + A_x) \sigma^y - (\hat{p}_y + A_y) \sigma^x) \psi^{\vphantom\dagger}.
\label{eq:rashba}
\end{equation}
Here $\hat{p}_\mu$ is the i-th component of the momentum operator, and
$\alpha_{\rm so}$ is the strength of the Rashba coupling. The dependence on
the minimal combination $\hat{{\bf p}} + {\bf A}$ is required by
  the gauge invariance of the action in Eq.~\eqref{eq:1}. Note that the
magnetic field continues to couple to $\sigma^z$. 

In the fixed Coulomb gauge, the momentum and gauge terms within the
Rashba anisotropy of Eq.~\eqref{eq:rashba} have distinct effects.  The
former, momentum term, may be considered at the mean field level, as an intrinsic spin
splitting in the spinon dispersion.  Taking this
into account, the ESR signal arises from vertical inter-band
transitions  \cite{Glenn2012}.   The variation of these transitions with momentum leads
to an intrinsic lineshape, from which  useful
information about van Hove and other special points of the spinon
bands may be extracted by a detailed analysis\cite{Luo2018}.
In Fermi liquids, this physics is responsible for chiral spin
resonance \cite{Finkelstein2005,Mishchenko2006}. In one dimensional
spin chains with {\em uniform} DM interaction, the same basic physics
leads to a splitting of the ESR line into a {\em doublet}
\cite{Povarov2011}.

The gauge field part of Eq.~\eqref{eq:rashba} consists mathematically of coupling of ${\bf A}$
to the spin-non-conserving bilinear operators of spinons:
\begin{equation}
  \label{eq:53}
H'_{\bm{A}}\! = \! \frac{i \alpha_{\rm so}}{2} \! \sum_{p,q}
[\psi^\dagger_{p+q, \downarrow} \psi_{p, \uparrow} A^+(q) \! -\! 
\psi^\dagger_{p+q, \uparrow} \psi_{p, \downarrow} A^-(q)]\!,
\end{equation}
where
$A^\pm = A_x \pm i A_y$.    This term has no
simple mean field description, and is responsible
for the magnetic field and temperature dependence of the dynamical
spin susceptibility, and in particular the ESR line width
$\eta$. 

Instead of a technically-involved diagrammatic calculation (which is
also possible and confirms the results otherwise obtained) we employ an
elegant short-cut which is based on the modern reincarnation
\cite{Oshikawa2002,Furuya2017} of the classic ESR formulation
by Mori and Kawasaki \cite{Mori1962}.   We are interested in the retarded Green's function of the transverse spin
fluctuations $G^R_{S^+S^{-}}(\bm{q}=0,\omega) = 2 M/(\omega - 2\omega_B - \Sigma(\omega))$,
which defines the zero momentum self-energy $\Sigma(\omega)$.  The ESR
theory shows (see \cite{SM}) that this self-energy is related to the correlations of
the perturbation operator ${\cal R} = [H'_{\bm{A}}, S^+]$, according
to
\begin{equation}
  \label{eq:55}
  \eta = \Im \Sigma(\omega) = -\frac{1}{2 M} \Im G^R_{{\cal R}{\cal
    R}^\dagger}(\omega).
\end{equation}
Eq.~\eqref{eq:55} directly expresses the ESR linewidth in terms of the 
retarded Green's function of the perturbing operator ${\cal R}$.
Observe that ${\cal R} \propto \alpha_{\rm so}$ and
hence to the second order in the spin-orbit coupling, Eq.~\eqref{eq:55}
may be calculated with respect to the isotropic Hamiltonian of the
ideal spin liquid subject to the Zeeman magnetic field $\omega_B$. 

\begin{figure}[h]
       \includegraphics[width=5.5cm]{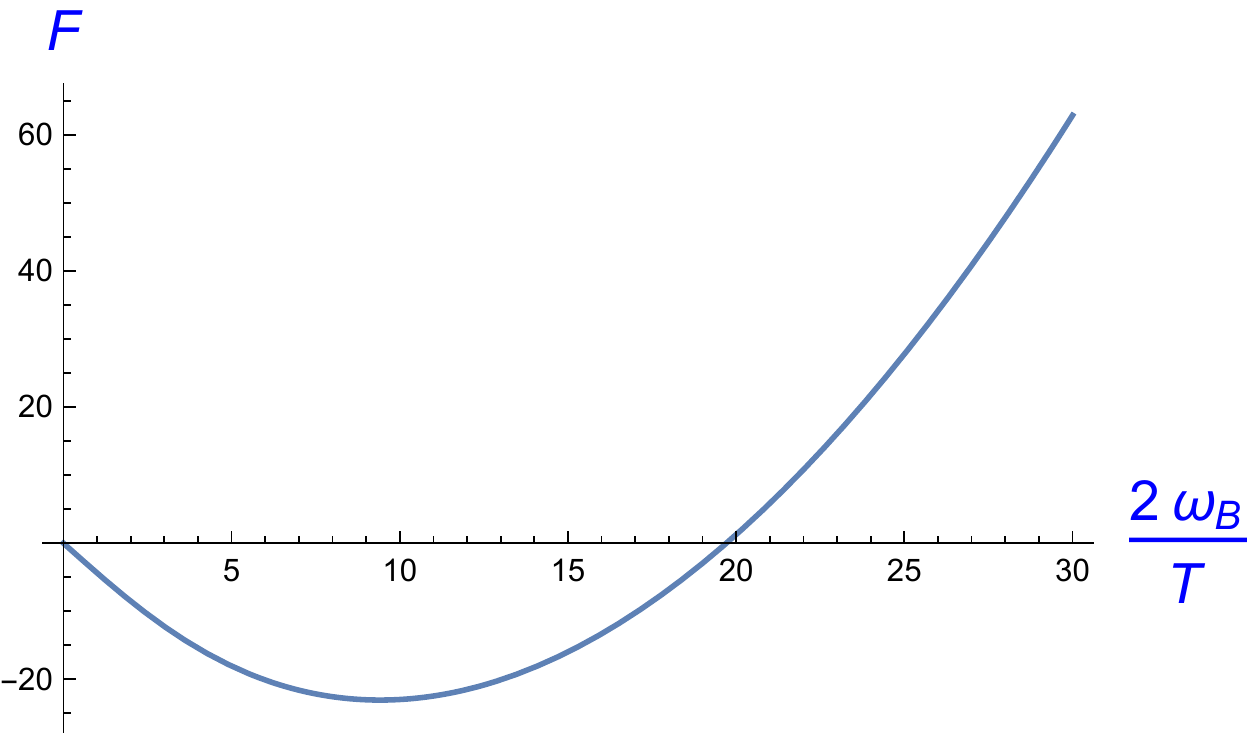}
    \caption{Scaling function $F(x)$.}
     \label{fig:eta}
\end{figure}

For the Rashba coupling, one obtains ${\cal R} = - i \frac{\alpha_{\rm so}}{2}
\sum_{p,q} \psi^\dagger_{p+q} \sigma^z\psi^{\vphantom\dagger}_{p} A^-(q)$, so that the calculation of $\eta$ reduces to a convolution-type 
integral over energy and momentum of the spectral functions of the spinon magnetization density $S^z({\bf q})$ and the gauge field ${\bm A}({\bm q})$. 
This instructive calculation is described in \cite{SM} and results in the
full scaling function prediction for the ESR linewidth
\begin{equation}
  \label{eq:56}
  \eta(\omega_B, T) = \frac{\alpha_{\rm so}^2}{2M} (\frac{m T}{8\pi \chi} + {\rm const}~ T^{5/3} F(\frac{2\omega_B}{T})).
\end{equation}
The scaling function $F(x)$ is plotted in Figure~\ref{fig:eta} and is
characterized by these limits: $F(x\ll 1) = -4.4 x$ and
$F(x \gg 1) = 0.75 x^{5/3}$.  Consequently, in the low-temperature
limit the linewidth follows a `fractional' scaling with the 
magnetic field,
$\eta \to \alpha_{\rm so}^2(2\omega_B)^{5/3}/M \sim \alpha_{\rm so}^2 B^{2/3}$. 
Also notable is the non-monotonic
dependence of the scaling function $F$ on its argument. The full
scaling function represents a non-trivial quantitative prediction for
the the present model of magnetic anisotropy.  

However, while all isotropic magnets are alike, all anisotropic magnets are anisotropic in their own way. 
We leave an exhaustive study of different mechanisms of anisotropy on
ESR in spin liquids for future work. 

\acknowledgements

We thank A. Furusaki, D. Golosov, E. Mishchenko, D. Maslov and M. Raikh for 
discussions. Our work is supported by the NSF CMMT program under
Grants No. DMR1818533 (L.B.) and DMR1928919 (O.A.S.), and we
benefitted from the facilities of the KITP, NSF grant PHY1748958.

\bibliography{spinons-ref.bib}
\widetext
\newpage 
\pagebreak

\begin{center}
\textbf{\large Supplemental Materials for ``Spinon waves in magnetized spin liquids"}\\
{\large Leon Balents and Oleg A. Starykh}
\end{center}
\setcounter{equation}{0}
\setcounter{figure}{0}
\setcounter{table}{0}
\setcounter{page}{1}
\makeatletter
\renewcommand{\theequation}{S\arabic{equation}}
\renewcommand{\thefigure}{S\arabic{figure}}
\renewcommand{\thesection}{\Alph{section}}

\section{Gauge field corrections to the transverse spin susceptibility}
\label{sec:gauge-field}

In the main text, we show that coupling to the gauge field induces spectral weight outside the region of the particle-hole continuum of the free fermion theory.  On physical grounds, this is expected because in addition to the fermionic ``quasiparticles'' (we use quotes because they are not gauge invariant and have a non-Fermi liquid self-energy) the system possesses collective gauge excitations with a very soft dispersion relation $\omega \sim k^3$.  By creating a particle-hole pair {\em and} a photon, one may shunt enough of the total momentum of the excitation in to the photon to bring the remainder into the kinematically allowed region for particle-hole pairs, and because the energy of the photon is so small, this should be possible at just about any energy.  With this picture in mind, we seek contributions to the dynamical spin structure factor with a single gauge field propagator, because ``cutting'' this line corresponds to a single excited photon excitation.  With a single gauge propagator, there are three diagrams, as shown in Eq.\eqref{eq:9}.  In the first and second contributions, $\chi^{1}_1(q,\omega_n)$ and $\chi^{1}_2(q,\omega_n)$, the gauge line does not cross the particle-hole bubble, so the gauge field acts here as a self-energy correction to one of the two fermion lines.  In the third diagram, the gauge line crosses the bubble, which represents a vertex correction.  Care must be taken to combine all three terms because, as shown by Kim {\em et al.}\cite{Kim1994}, there are important cancellations between them which are required to maintain gauge invariance and avoid unphysical results at low frequency and momentum.

The formal expressions for these three contributions are
\begin{align}
  \label{eq:10}
  \chi^{1}_1(q,\omega_n) & = \int (dk) (dp) G_\uparrow^2(\bm{k},k_n) G_\uparrow(\bm{k}+\bm{p},k_n+p_n) G_\downarrow(\bm{k}+\bm{q},k_n+\omega_n) \frac{2k_\mu + p_\mu}{2m} \frac{2k_\nu+p_\nu}{2m} D_{\mu\nu}(\bm{p},p_n),\nonumber \\
  \chi^{1}_2(q,\omega_n) & = \int (dk) (dp) G_\uparrow(\bm{k},k_n) G_\downarrow^2(\bm{k}+\bm{q},k_n+\omega_n) G_\downarrow(\bm{k}+\bm{p}+\bm{q},k_n+p_n+\omega_n) \nonumber \\
  & \hspace{1.0in} \times\frac{2k_\mu + 2q_\mu+p_\mu}{2m} \frac{2k_\nu+2q_\nu+p_\nu}{2m} D_{\mu\nu}(\bm{p},p_n), \nonumber \\
  \chi^{1}_3(q,\omega_n) & = \int (dk) (dp) G_\uparrow(\bm{k},k_n) G_\uparrow(\bm{k}+\bm{p},k_n+p_n) G_\downarrow(\bm{k}+\bm{q},k_n+\omega_n) G_\downarrow(\bm{k}+\bm{p}+\bm{q},k_n+p_n+\omega_n)  \nonumber \\
  & \hspace{1.0in} \times \frac{2k_\mu + p_\mu}{2m} \frac{2k_\nu+2q_\nu+p_\nu}{2m} D_{\mu\nu}(\bm{p},p_n),
\end{align}
where we introduced bold face for spatial vectors, $k_n$ and $p_n$ as Matsubara frequencies, and the notation $(dk) = d^2\bm{k} dk_n /(2\pi)^3$.   The gauge propagator is
\begin{equation}
  \label{eq:11}
  D_{\mu\nu}(\bm{p},p_n) = \left(\delta_{\mu\nu} - \frac{p_\mu p_\nu}{p^2}\right) D(\bm{p},p_n),
\end{equation}
with
\begin{equation}
  \label{eq:12}
  D(\bm{p},p_n) = \frac{1}{\gamma |p_n|/p + \chi p^2}.
\end{equation}
Here in a strict $1/N$ expansion the $\gamma$ coefficient, which reflects Landau damping, itself originates from a fermion bubble.  We will treat it however as just a bare kinetic term for the gauge field, following numerous previous works.

Consider the first two diagrams.  In their expressions, the integral over $p$ defines the self-energy $\Sigma$, and these terms can be rewritten in terms of $\Sigma$.  Specifically,
\begin{align}
  \label{eq:13}
  \chi^{1}_1(q,\omega_n) & = \int (dk) \, \Sigma_\uparrow(\bm{k},k_n) G_\uparrow^2(\bm{k},k_n) G_\downarrow(\bm{k}+\bm{q},k_n+\omega_n),
                      \nonumber \\
    \chi^{1}_2(q,\omega_n) & = \int (dk) \, \Sigma_\downarrow(\bm{k}+\bm{q},k_n+\omega_n) G_\uparrow(\bm{k},k_n) G^2_\downarrow(\bm{k}+\bm{q},k_n+\omega_n),  
\end{align}
where the self-energy is
\begin{align}
  \label{eq:14}
  \Sigma_\alpha(\bm{k},k_n) & = \int (dp) G_\alpha(\bm{k}+\bm{p},k_n+p_n) \frac{2k_\mu + p_\mu}{2m} \frac{2k_\nu+p_\nu}{2m} D_{\mu\nu}(\bm{p},p_n) \nonumber \\
  & = \int (dp) \left(\frac{\bm{k}\times \hat{\bm{p}}}{m}\right)^2 D(\bm{p},p_n) G_\alpha(\bm{k}+\bm{p},k_n+p_n) .
\end{align}
The self-energy is a standard calculation in the spinon gauge theory.  Writing the Green's function explicitly, we have
\begin{equation}
  \label{eq:15}
  \Sigma_\alpha(\bm{k},k_n) = \int (dp) \left(\frac{\bm{k}\times \hat{\bm{p}}}{m}\right)^2 \frac{D(\bm{p},p_n)}{ik_n+ip_n- \epsilon_{\bm{k}+\bm{p},\alpha}} ,
\end{equation}
with
\begin{align}
  \label{eq:23}
  \epsilon_{\bm{k},\alpha} = \xi_k -\alpha \omega_B + U n_{-\alpha},
\end{align}
with $\alpha=\uparrow=+1$ and $\alpha=\downarrow = -1$, and $\xi_k = (k^2-k_F^2)/2m$.
Owing to the singular nature of the gauge propagator, the integral for the self-energy is dominated by small $\bm{p}$.  Choosing coordinates $\bm{p} = \bm{\hat{k}} p_\parallel + \bm{\hat{z}}\times \bm{\hat{k}} p_\perp$, we have
\begin{align}
  \label{eq:16}
  \epsilon_{\bm{k}+\bm{p},\alpha} \approx \epsilon_{\bm{k},\alpha} + v_F p_\parallel  + \frac{p_\perp^2}{2m}.
\end{align}
Here $ \epsilon_{\bm{k},\alpha}$ includes a spin-dependent Zeeman shift and a Hartree correction, which are both constant.  From this form we expect the scaling $p_\parallel \sim p_\perp^2 \ll p_\perp$, which means that $\bm{p}$ is approximately normal to $\bm{k}$.  This means we can replace $p$ by $p_\perp$ inside the gauge propagator, and that $|\bm{k}\times \hat{\bm{p}}|^2 =k^2 p_\perp^2 \approx k^2 \approx k_F^2$, for momentum $\bm{k}$ near the Fermi surface.  Consequently, we have
\begin{align}
  \label{eq:17}
  \Sigma_\alpha(\bm{k},k_n) \approx \frac{k_F^2}{m^2} \int \frac{dp_n}{2\pi} \frac{dp_\perp}{2\pi} \frac{dp_\parallel}{2\pi} \left( \frac{|p_\perp|}{\gamma |p_n| + \chi |p_\perp|^3}\right) \frac{1}{ik_n+ip_n -\epsilon_{\bm{k},\alpha} - p_\perp^2/2m - v_F p_\parallel}.
\end{align}
The integral over $p_\parallel$ can be done immediately -- there is a small subtlety in the real part of the integral is conditionally convergent and dependent upon the cutoff, but this is anyway absorbed in a simple Fermi energy shift and can be set to zero.  The imaginary part is well-defined and we obtain
\begin{align}
  \label{eq:18}
  \Sigma_\alpha(\bm{k},k_n) = \frac{-i k_F^2}{2m^2 v_F}  \int \frac{dp_n}{2\pi} \frac{dp_\perp}{2\pi} \frac{|p_\perp| {\rm sign}(k_n+p_n)}{\gamma |p_n| + \chi |p_\perp|^3}.
\end{align}
Note that the dependence on momentum $\bm{k}$ has dropped out, so the self-energy is purely frequency dependent -- what is called a ``local'' self-energy.  The actual function can be calculated by performing the $p_n$ integration first: the regions at large positive and negative frequency cancel one another and the full result is just obtained from the integral between $-|k_n|$ and $|k_n|$.  We find
\begin{align}
  \label{eq:32}
  \Sigma_\alpha(\bm{k},k_n) = \frac{-i v_F}{2\pi\gamma} {\rm sign} (k_n) \int \! \frac{dp_\perp}{2\pi}\, |p_\perp| \ln \frac{\gamma|k_n| + \chi |p_\perp|^3}{\chi |p_\perp|^3}.
\end{align}
Then carrying out the $p_\perp$ integration gives finally
\begin{equation}
  \label{eq:19}
  \Sigma_\alpha(\bm{k},k_n) = \frac{-i v_F}{2\sqrt{3} \pi }\frac{{\rm sign}(k_n)|k_n|^{2/3}}{\gamma^{1/3} \chi^{2/3}}.
\end{equation}
The 2/3 power law dependence on frequency of the self-energy is a famous result for the spinon Fermi surface.  Now we will rewrite Eqs.~\eqref{eq:13} in a form more amenable to seeing the partial cancellations of the three diagrams.  To do so, we use the partial fractions rewriting
\begin{align}
  \label{eq:20}
  G_\uparrow^2(\bm{k},k_n) G_\downarrow(\bm{k}+\bm{q},k_n+\omega_n) & = \frac{G_\uparrow(\bm{k},k_n) \left[G_\uparrow(\bm{k},k_n)-G_\downarrow(\bm{k}+\bm{q},k_n+\omega_m)\right]}{i\omega_n - \epsilon_{\bm{k}+\bm{q},\downarrow} + \epsilon_{\bm{k},\uparrow}}, \\
  G_\downarrow(\bm{k}+\bm{q},k_n+\omega_n)^2 G_\uparrow(\bm{k},k_n) & =  \frac{G_\downarrow(\bm{k}+\bm{q},k_n+\omega_n) \left[G_\downarrow(\bm{k}+\bm{q},k_n+\omega_m)-G_\uparrow(\bm{k},k_n)\right]}{-i\omega_n + \epsilon_{\bm{k}+\bm{q},\downarrow} - \epsilon_{\bm{k},\uparrow}},\nonumber
\end{align}
which is straightforward to show using of the explicit forms for the Green's functions.  Using these forms, we obtain the sum of the two self-energy contributions as $\chi^{1}_{12} = \chi^{1}_1+\chi^{1}_2$
\begin{align}
  \label{eq:21}
  \chi^{1}_{12}(\bm{q},\omega_n) & = \int (dk) \frac{\Sigma_\downarrow(\bm{k}+\bm{q},k_n+\omega_n) - \Sigma_\uparrow(\bm{k},k_n)}{i\omega_n+\epsilon_{\bm{k}\uparrow}-\epsilon_{\bm{k}+\bm{q},\downarrow}} G_\uparrow(\bm{k},k_n) G_\downarrow(\bm{k}+\bm{q},k_n+\omega_n) \nonumber \\
  & + \int (dk) \frac{\Sigma_\uparrow(\bm{k},k_n) G^2_\uparrow(\bm{k},k_n)-\Sigma_\downarrow(\bm{k}+\bm{q},k_n+\omega_n) G^2_\downarrow(\bm{k}+\bm{q},k_n+\omega_n) }{i\omega_n +\epsilon_{\bm{k}\uparrow}-\epsilon_{\bm{k}+\bm{q},\downarrow}} .
\end{align}

We aim to massage the susceptibility corrections into a form which exposes the small $q$ dependence more clearly and makes physical interpretation easier.  
First we split \eqref{eq:21} into two distinct parts,
\begin{equation}
  \label{eq:o1}
  \chi^{1}_{12} = \chi^{1}_{12,A}+\chi^{1}_{12,B},
\end{equation}
with
\begin{align}
  \chi^{1}_{12,A}(\bm{q},\omega_n) & = \int (dk) \frac{\Sigma_\downarrow(\bm{k}+\bm{q},k_n+\omega_n) - \Sigma_\uparrow(\bm{k},k_n)}{i\omega_n+\epsilon_{\bm{k}\uparrow}-\epsilon_{\bm{k}+\bm{q},\downarrow}} G_\uparrow(\bm{k},k_n) G_\downarrow(\bm{k}+\bm{q},k_n+\omega_n)  \\
  \chi^{1}_{12,B}(\bm{q},\omega_n) & =  \int (dk) \frac{\Sigma_\uparrow(\bm{k},k_n) G^2_\uparrow(\bm{k},k_n)-\Sigma_\downarrow(\bm{k}+\bm{q},k_n+\omega_n) G^2_\downarrow(\bm{k}+\bm{q},k_n+\omega_n) }{i\omega_n +\epsilon_{\bm{k}\uparrow}-\epsilon_{\bm{k}+\bm{q},\downarrow}} .
  \label{eq:o2}
\end{align}
We show below that $\chi^{1,B}_{12}$ does not contribute to the imaginary part of the susceptibility and for a moment just neglect it.  Next we use the identity
\begin{align}
  G_\uparrow(k) G_\downarrow(k+q) & = \frac{G_\uparrow(k)-G_\downarrow(k+q)}{i\omega_n + \epsilon_\uparrow(\bm{k}) - \epsilon_\downarrow(\bm{k}+\bm{q})} 
  \label{eq:o3}
\end{align}
to write
\begin{align}
   \chi^{1}_{12,A}(\bm{q},\omega_n) & = \int (dk) \frac{\Sigma_\downarrow(\bm{k}+\bm{q},k_n+\omega_n) - \Sigma_\uparrow(\bm{k},k_n)}{(i\omega_n+\epsilon_{\bm{k}\uparrow}-\epsilon_{\bm{k}+\bm{q},\downarrow})^2}\left[G_\uparrow(k)-G_\downarrow(k+q)\right].
   \label{eq:o4}
\end{align}
Next we use the expression for the self-energy, \eqref{eq:14}, and obtain
\begin{align}
   \chi^{1}_{12,A}(\bm{q},\omega_n) & = \frac{1}{m^2}\int (dk)(dp) \frac{D(\bm{p},p_n)}{(i\omega_n+\epsilon_{\bm{k}\uparrow}-\epsilon_{\bm{k}+\bm{q},\downarrow})^2}\left[ \left((\bm{k}+\bm{q})\times \bm{\hat{p}}\right)^2 G_\downarrow(k+q+p) - \left(\bm{k}\times\bm{\hat{p}}\right)^2 G_\uparrow(k+p)\right]\left[G_\uparrow(k)-G_\downarrow(k+q)\right].
   \label{eq:o5}
\end{align}
Similar manipulations of the vertex diagram give
\begin{align}
  \chi^{1}_{3}(\bm{q},\omega_n)  =\frac{1}{m^2}\int (dk) (dp) & \frac{D(\bm{p},p_n)}{(i\omega_n+\epsilon_{\bm{k}\uparrow}-\epsilon_{\bm{k}+\bm{q},\downarrow}) (i\omega_n+\epsilon_{\bm{k}+\bm{p},\uparrow}-\epsilon_{\bm{k}+\bm{p}+\bm{q},\downarrow})} (\bm{k}\times\bm{\hat{p}})\cdot((\bm{k}+\bm{q})\times\bm{\hat{p}})\nonumber \\
  &\times \left[ G_\uparrow(k+p) - G_\downarrow(k+q+p)  \right]\left[G_\uparrow(k)-G_\downarrow(k+q)\right].  
  \label{eq:o7}
\end{align}

\section{Imaginary part}
\label{sec:imaginary}

Let us consider the imaginary part of the (real frequency)
susceptibility {\em outside} the particle hole (PH) continuum.  This means
that the explicit denominators in Eqs.~(\ref{eq:o5},\ref{eq:o7}) can be
trivially analytically continued and considered purely real.  We will
however have to do the $p_n$ and $k_n$ integrals.  

Start with \eqref{eq:o5} and observe that if the product of the two
square brackets is multiplied out, two of the four resulting terms,
those involving the integral of $G_\downarrow(k+p+q)
G_\downarrow(k+q)$ and $G_\uparrow(k+p) G_\uparrow(k)$, do not have
imaginary part outside the PH continuum simply because the integral
over $k$ gives a result which is
$\omega_n$-independent (in the first of these we can make a shift
$k\to k-q$  and then integrate over $k$). That is, the only source of
the imaginary part in these contributions is provided by the denominator in 
$1/(i\omega_n+\epsilon_{\bm{k}\uparrow}-\epsilon_{\bm{k}+\bm{q},\downarrow})$ -- and its imaginary part is restricted to the non-interacting PH continuum. The same argument actually applies to 
$\chi^{1}_{12,B}$ in \eqref{eq:o2}, one just needs to shift $k\to k-q$ in the 2nd term on the right hand side there.

The two terms in $\chi^{1}_{12,A}$ that contribute outside the PH continuum are therefore given by
\begin{align}
   \chi^{1}_{12,A}(\bm{q},\omega_n) & = \frac{1}{m^2}\int (dk)(dp) \frac{D(\bm{p},p_n)}{(i\omega_n+\epsilon_{\bm{k}\uparrow}-\epsilon_{\bm{k}+\bm{q},\downarrow})^2}
   \left[ \left((\bm{k}+\bm{q})\times \bm{\hat{p}}\right)^2 G_\downarrow(k+q+p)G_\uparrow(k) + \left(\bm{k}\times\bm{\hat{p}}\right)^2 G_\uparrow(k+p) G_\downarrow(k+q)\right].  
    \label{eq:o8}
\end{align}
Since the gauge propagator \eqref{eq:12} is even in $p=(\bf{p}, p_n)$, we can transform $p\to -p$, followed by $k\to k+p$, in the first term in the square brackets above to obtain
\begin{align}
   \chi^{1}_{12,A}(\bm{q},\omega_n)  = \frac{1}{m^2}\int (dk)(dp) D(\bm{p},p_n) \left[ \frac{(\bm{k}\times \bm{\hat{p}})^2}{(i\omega_n+\epsilon_{\bm{k},\uparrow}-\epsilon_{\bm{k}+\bm{q},\downarrow})^2}
   + \frac{((\bm{k}+\bm{q})\times \bm{\hat{p}})^2}{(i\omega_n+\epsilon_{\bm{k}+\bm{p}, \uparrow}-\epsilon_{\bm{k}+\bm{q} + \bm{p} ,\downarrow})^2} \right]
   G_\uparrow(k+p) G_\downarrow(k+q).  
\label{eq:o9}
\end{align}
Now insert the spectral representation,
\begin{equation}
  \label{eq:54}
  D(\bm{p},p_n) = \int \! d\nu \, \frac{d(\bm{p},\nu)}{ip_n-\nu}.
\end{equation}
Here we want
\begin{equation}
  \label{eq:80}
  D(\bm{p},p_n) = \frac{1}{\gamma |p_n|/p+\chi p^2}.
\end{equation}
In the usual way, we extract the spectral function via
\begin{equation}
  \label{eq:81}
  d(\bm{p},\omega) =-\frac{1}{\pi} {\rm Im} D(\bm{p},ip_n \rightarrow
  \omega+i0^+) = -\frac{1}{\pi} \frac {\gamma\omega p}{\gamma^2
    \omega^2 + \chi^2 p^6}.
\end{equation}
Eq.\eqref{eq:o9} factorizes into
\begin{align}
   \chi^{1}_{12,A}(\bm{q},\omega_n)  = \frac{1}{m^2}\int (d\bm{k})(d\bm{p}) \left[ \frac{(\bm{k}\times \bm{\hat{p}})^2}{(i\omega_n+\epsilon_{\bm{k},\uparrow}-\epsilon_{\bm{k}+\bm{q},\downarrow})^2}
   + \frac{((\bm{k}+\bm{q})\times \bm{\hat{p}})^2}{(i\omega_n+\epsilon_{\bm{k}+\bm{p}, \uparrow}-\epsilon_{\bm{k}+\bm{q} + \bm{p} ,\downarrow})^2} \right] I_{12},
 \label{eq:o10}
\end{align}
where 
\begin{align}
I_{12} = \int d\nu d(\bm{p},\nu) \int \frac{dk_n dp_n}{(2\pi)^2} \frac{1}{i p_n - \nu} \frac{1}{i k_n + i p_n - \epsilon_{\bm{k}+\bm{p}, \uparrow}} \frac{1}{i k_n + i \omega_n - \epsilon_{\bm{k}+\bm{q}, \downarrow}}.
\label{eq:o11}
\end{align}
Carrying out the contour integrations we obtain
\begin{align}
I_{12} = -\int d\nu \frac{d(\bm{p},\nu) }{i\omega_n + \epsilon_{\bm{k}+\bm{p}, \uparrow} - \epsilon_{\bm{k}+\bm{q}, \downarrow} - \nu}\left[\theta(\nu)\theta(-\epsilon_{\bm{k}+\bm{p}, \uparrow}) 
\theta( \epsilon_{\bm{k}+\bm{q}, \downarrow}) + \theta(-\nu) \theta(\epsilon_{\bm{k}+\bm{p}, \uparrow}) \theta( -\epsilon_{\bm{k}+\bm{q}, \downarrow})\right].
\label{eq:o12}
\end{align}
Now we can analytically continue $i\omega_n \to \omega + i0$, with
$\omega > 0$, and obtain the imaginary part, which constraints $\nu = \omega + \epsilon_{\bm{k}+\bm{p}, \uparrow} - \epsilon_{\bm{k}+\bm{q}, \downarrow}$ and collapses the $\nu$-integration
\begin{align}
{\rm Im} I_{12} = \pi d(\bm{p}, \omega + \epsilon_{\bm{k}+\bm{p}, \uparrow} - \epsilon_{\bm{k}+\bm{q}, \downarrow}) \theta( \epsilon_{\bm{k}+\bm{q}, \downarrow}) \theta(-\epsilon_{\bm{k}+\bm{p}, \uparrow})
\theta( \omega + \epsilon_{\bm{k}+\bm{p}, \uparrow} - \epsilon_{\bm{k}+\bm{q}, \downarrow}) ~{\rm for} ~\omega > 0.
\label{eq:o13}
\end{align}
We observe that the second set of step-functions in \eqref{eq:o12} is
always zero for $\omega > 0$.

Now consider the
vertex part, Eq.~\eqref{eq:o7}.  Expanding the product in the
second line, we seek terms that have some $\omega_n=q_n$
dependence.  There are just two such parts
\begin{align}
  \chi^{1}_{3}(\bm{q},\omega_n)  = -\frac{1}{m^2}\int (dk) (dp) & \frac{D(\bm{p},p_n)}{(i\omega_n+\epsilon_{\bm{k}\uparrow}-\epsilon_{\bm{k}+\bm{q},\downarrow}) (i\omega_n+\epsilon_{\bm{k}+\bm{p},\uparrow}-\epsilon_{\bm{k}+\bm{p}+\bm{q},\downarrow})} (\bm{k}\times\bm{\hat{p}})\cdot((\bm{k}+\bm{q})\times\bm{\hat{p}})\nonumber \\
  & \times \left[ G_\uparrow(k+p) G_\downarrow(k+q) + G_\downarrow(k+q+p) G_\uparrow(k) \right].  
  \label{eq:o14}
\end{align}
We again observe that the set of transformations $p\to -p$, followed
by $k\to k+p$, make the 2nd term in the square brackets above
equal to the 1st, so that
\begin{align}
  \chi^{1}_{3}(\bm{q},\omega_n)  = -\frac{2}{m^2}\int (dk) (dp) & \frac{(\bm{k}\times\bm{\hat{p}})\cdot((\bm{k}+\bm{q})\times\bm{\hat{p}})}{(i\omega_n+\epsilon_{\bm{k}\uparrow}-\epsilon_{\bm{k}+\bm{q},\downarrow}) (i\omega_n+\epsilon_{\bm{k}+\bm{p},\uparrow}-\epsilon_{\bm{k}+\bm{p}+\bm{q},\downarrow})} D(\bm{p},p_n)  G_\uparrow(k+p) G_\downarrow(k+q).
  \label{eq:o15}
\end{align}
Evidently it reduces to a form very similar to \eqref{eq:o10}. Namely,
\begin{align}
  \chi^{1}_{3}(\bm{q},\omega_n)  = -\frac{2}{m^2}\int (d\bm{k}) (d\bm{p}) & \frac{(\bm{k}\times\bm{\hat{p}})\cdot((\bm{k}+\bm{q})\times\bm{\hat{p}})}{(i\omega_n+\epsilon_{\bm{k}\uparrow}-\epsilon_{\bm{k}+\bm{q},\downarrow}) (i\omega_n+\epsilon_{\bm{k}+\bm{p},\uparrow}-\epsilon_{\bm{k}+\bm{p}+\bm{q},\downarrow})} I_{12}.
  \label{eq:o16}
\end{align}
Therefore, the imaginary part of $\chi^1 = \chi^1_{12} + \chi^1_3$ is determined by the imaginary part of $I_{12}$ \eqref{eq:o13},
\begin{align}
  {\rm Im} \chi^{1}(\bm{q},\omega) & = \frac{1}{m^2}\int (d\bm{k})(d\bm{p}) \Big[ \frac{(\bm{k}\times \bm{\hat{p}})^2}{(\omega+\epsilon_{\bm{k},\uparrow}-\epsilon_{\bm{k}+\bm{q},\downarrow})^2}
   + \frac{((\bm{k}+\bm{q})\times \bm{\hat{p}})^2}{(\omega+\epsilon_{\bm{k}+\bm{p}, \uparrow}-\epsilon_{\bm{k}+\bm{q} + \bm{p} ,\downarrow})^2} \nonumber\\
  & - \frac{2(\bm{k}\times\bm{\hat{p}})\cdot((\bm{k}+\bm{q})\times\bm{\hat{p}})}{(\omega+\epsilon_{\bm{k}\uparrow}-\epsilon_{\bm{k}+\bm{q},\downarrow}) 
  (\omega+\epsilon_{\bm{k}+\bm{p},\uparrow}-\epsilon_{\bm{k}+\bm{p}+\bm{q},\downarrow})} \Big] {\rm Im} I_{12}.
 \label{eq:o17}
\end{align}
At this point, we can see that the quantity in the square brackets in
Eq.~\eqref{eq:o17} vanishes at $q=0$.  Hence we may expect a quadratic dependence on $q$ (although this quantity has linear terms in $\bm{q}$,  they vanish on integration or pair off with another linear part from $I_{12}$).

Let us see how to make the integration explicit.  We choose polar coordinates according to
\begin{align}
  \label{eq:76}
  \bm{q} & = (q,0), \nonumber \\
  \bm{k} & = \bm{k}_F + \delta \bm{k} = (k_F + \delta k) (\cos\theta,\sin\theta), \nonumber\\
  \bm{p} & = \bm{p}_\parallel + \bm{p}_\perp = p_\parallel (\cos\theta,\sin\theta) + p_\perp (-\sin\theta,\cos\theta).
\end{align}
We assume $\delta k, q, p_\parallel,p_\perp \ll k_F$.  The vector
vertices in \eqref{eq:o17} can be simplified as follows:
\begin{align}
\bm{k}\times \bm{\hat{p}} &= \hat{z} (k_x p_y - k_y p_x)/p = \hat{z} \frac{k_F + \delta k}{p} [\cos\theta (p_\parallel \sin\theta + p_\perp \cos\theta) - \sin\theta (p_\parallel \cos\theta - p_\perp \sin\theta)] = 
\hat{z} \frac{k_F + \delta k}{p} p_\perp \nonumber\\
& \approx \hat{z} k_F \frac{p_\perp}{p} .
\label{eq:o18}
\end{align}
At this point we need to compare $p_\parallel$ and $p_\perp$. From $\epsilon_{\bm{k}+\bm{p}} - \epsilon_{\bm{k}} = v_F p_\parallel + \frac{p_\perp^2}{2m} + \frac{p_\parallel^2}{2m}$ we deduce that
$p_\parallel \sim p_\perp^2/(m v_F) = p_\perp^2/k_F \ll p_\perp$. Therefore $\frac{p_\parallel^2}{2m}$ can be neglected in comparison with the two first terms and, moreover, 
$p = \sqrt{p_\perp^2 + p_\parallel^2} \approx p_\perp$. Hence \eqref{eq:o18} reduces to just $\hat{z} k_F$. 

At the same time we see that $\bm{q}\times \bm{\hat{p}} \approx \hat{z} q \sin\theta \ll \hat{z} k_F$. Therefore all vertices in \eqref{eq:o17} can be safely approximated by $\hat{z} k_F$.
This strongly simplifies \eqref{eq:o17},
\begin{align}
  {\rm Im} \chi^{1}(\bm{q},\omega) & = \frac{k_F^2}{m^2}\int (d\bm{k})(d\bm{p}) \Big[ \frac{1}{(\omega+\epsilon_{\bm{k},\uparrow}-\epsilon_{\bm{k}+\bm{q},\downarrow})^2}
   + \frac{1}{(\omega+\epsilon_{\bm{k}+\bm{p}, \uparrow}-\epsilon_{\bm{k}+\bm{q} + \bm{p} ,\downarrow})^2} 
   - \frac{2}{(\omega+\epsilon_{\bm{k}\uparrow}-\epsilon_{\bm{k}+\bm{q},\downarrow}) 
  (\omega+\epsilon_{\bm{k}+\bm{p},\uparrow}-\epsilon_{\bm{k}+\bm{p}+\bm{q},\downarrow})} \Big] \nonumber\\ 
  &\times {\rm Im} I_{12} 
  = v_F^2 \int (d\bm{k})(d\bm{p}) \Big[ \frac{1}{\omega+\epsilon_{\bm{k},\uparrow}-\epsilon_{\bm{k}+\bm{q},\downarrow}} - 
  \frac{1}{\omega+\epsilon_{\bm{k}+\bm{p}, \uparrow}-\epsilon_{\bm{k}+\bm{q} + \bm{p} ,\downarrow}}\Big]^2 {\rm Im} I_{12}.
 \label{eq:o19}
\end{align}
Next we observe that $\epsilon_{\bm{k}+\bm{p}, \uparrow} +
\epsilon_{\bm{k}+\bm{q},\downarrow} - \epsilon_{\bm{k},\uparrow} -
\epsilon_{\bm{k}+\bm{q} + \bm{p} ,\downarrow} = -2 \bm{q} \cdot \bm{p}
\approx 2 q p_\perp \sin\theta$.  This combination appears when
combining denominators in the squared factor in Eq.~\eqref{eq:o19},
and explicitly demonstrates the $q^2$ dependence of the result.

Next we can write
\begin{align}
  \label{eq:77}
  \epsilon_{\bm{k}+\bm{q},\downarrow} & \approx v_F (\delta k + q\cos\theta) + \tilde{\omega}_B, \\
  \epsilon_{\bm{k}+\bm{p},\uparrow} & \approx v_F (\delta k + p_\parallel)+ p_\perp^2/(2m)- \tilde{\omega}_B, \\
  \epsilon_{\bm{k}+\bm{q} + \bm{p} ,\downarrow} & \approx v_F (\delta k +  p_\parallel +  q \cos\theta) + \frac{p_\perp^2}{2m} - \frac{p_\perp q \sin\theta}{m} + \tilde{\omega}_B.
\end{align}
The sign constraints on the energy in \eqref{eq:o13} therefore are
\begin{align}
  \label{eq:78}
  \epsilon_{\bm{k}+\bm{q},\downarrow}  >0 & \rightarrow v_F (\delta k +
                                            q\cos\theta) +\tilde{\omega}_B >0 , \\
  \epsilon_{\bm{k}+\bm{p},\uparrow}  <0 & \rightarrow v_F (\delta k +
                                          p_\parallel) +
                                          p _\perp^2/(2m) - \tilde{\omega}_B<0, \nonumber
 \end{align}
The frequency argument of the photon spectral function is
\begin{equation}
  \label{eq:80}
  \nu = \omega +\epsilon_{\bm{k}+\bm{p},\uparrow}-
  \epsilon_{\bm{k}+\bm{q},\downarrow} = \omega - 2\tilde{\omega}_B - v_F q \cos\theta
  + v_F p_\parallel + p_\perp^2/(2m).
\end{equation}
Eq.\eqref{eq:78} shows that $\nu$ is bounded by $\omega$ from above, $\nu \leq \omega$. At the same time $\nu > 0$.
Let us use Eq.~\eqref{eq:80} as a definition to trade the
$p_\parallel$ integration for one over $\nu$.  We have $dp_\parallel =
d\nu/v_F$.  In particular we see that
\begin{equation}
  \label{eq:79}
  v_F p_\parallel + p_\perp^2/(2m) = \nu+ 2\tilde{\omega}_B - \omega + v_F q \cos\theta.
\end{equation}
Introduce $\varepsilon$
\begin{align}
  \label{eq:84}
  \varepsilon = v_F \delta k + v_F q \cos\theta + \tilde{\omega}_B,
\end{align}
so that $d(\delta k) = d\varepsilon/v_F$. 

Hence the second line in Eq.~\eqref{eq:78} becomes $\varepsilon < \omega - \nu$ while the first line just reads $\varepsilon > 0$. Hence we obtain
\begin{equation}
  \label{eq:82}
 0 < \varepsilon < \omega - \nu.
\end{equation}

Now we can express all the energies in these variables.  
\begin{align}
  \label{eq:86}
   \epsilon_{\bm{k}+\bm{q},\downarrow} & \approx \varepsilon, \\
  \epsilon_{\bm{k}+\bm{p},\uparrow} &        \approx \nu-\omega  +\varepsilon , \\
  \epsilon_{\bm{k}\uparrow} & \approx v_F \delta k - \tilde{\omega}_B \approx
                              \epsilon-2h - v_F q \cos \theta, \\
  \epsilon_{\bm{k}+\bm{p}+\bm{q},\downarrow} & \approx v_F (\delta k +
                                               q\cos\theta +
                                               p_\parallel) +
                                               p_\perp^2/2m + \tilde{\omega}_B
                                               \approx \varepsilon+
                                               \nu - \omega+2\tilde{\omega}_B + v_F q \cos\theta.
\end{align}

Putting everything together and using $p \approx p_\perp$ and \eqref{eq:81} for $d(p_\perp,\nu)$, we obtain for \eqref{eq:o19}
\begin{align}
  {\rm Im} \chi^{1}(\bm{q},\omega) &
  = 4 v_F^2 q^2 \int (d\bm{k})(d\bm{p}) \frac{p_\perp^2\sin^2\theta}{(\omega - 2\tilde{\omega}_B - v_F q \cos\theta)^2 
(\omega - 2\tilde{\omega}_B - v_F q \cos\theta + q p_\perp
                                     \sin\theta/m)^2} {\rm Im} I_{12}
                                     \nonumber \\
& \approx - \frac{4 v_F^2 q^2}{(2\pi)^4} \int_{-\infty}^\infty dp_\perp \int_0^\omega \frac{d\nu}{v_F} \int_0^{2\pi} d\theta \int_0^{\omega-\nu} \frac{d\epsilon}{v_F} k_F 
\frac{p_\perp^2\sin^2\theta}{(\omega - 2\tilde{\omega}_B - v_F q \cos\theta)^4} \frac{\gamma \nu |p_\perp|}{\gamma^2 \nu^2 + \chi^2 p_\perp^6}.
 \label{eq:o20}
 \end{align}
 
Now we can finally complete the evaluation.  We
integrate over $p_\perp$ first, using $\int_0^\infty du u/(1+u^3) =
2\pi/3\sqrt{3}$, and obtain that $ {\rm Im} \chi^{1} \sim
\int_0^\omega d\nu (\omega-\nu)\nu^{1/3}$.   Then we obtain the final result
\begin{align}
  \label{eq:88}
  {\rm Im}  \, \chi^1(\bm{q},\omega) & \approx -\frac{\sqrt{3} \gamma^{1/3} k_F}{56\pi^2 \chi^{4/3}} \frac{q^2 \omega^{7/3}}{(\omega - 2\tilde{\omega}_B)^4}.
                                       \end{align}
The characteristic scaling $q^2 \omega^{7/3}$ is now explicitly shown.

\section{Real part}
\label{sec:real}

The calculations of the real part are more difficult. In accordance
with the common wisdom, see for example Chap.37 in the book by Abrikosov, Gor'kov,
Dzyaloshinskii, the 
outcome depends on the order of the integration over Matsubara
frequencies $k_n, p_n$ and momenta $\bm{k}, \bm{p}$.   Generally the
first integrals carried out can be calculated simply and exactly {\em
  in the absence of any ultraviolet cut-off}.   For a zero temperature
quantum system, there is indeed no frequency cut-off, so integrating
over all  frequencies is correct.  Integrating over all momenta is often not
correct, because there is some lattice or bandwidth scale.
Nevertheless, it can be tempting and simpler to try.  In our case, carrying out
the momentum integration first, similar to the procedure used by Kim
{\em et al.}\cite{Kim1994}, gives the incorrect result 
\begin{align}
  \label{eq:62}
\left. {\rm Re}\, \chi^1(\bm{q},\omega) \right|_{\textrm{
                                            $k$ integral first}}  & \simeq
                                            \frac{1}{8\pi^3\chi}
                                            \frac{v_F^2q^2
                                            \omega^2}{(\omega -
                                            2\tilde{\omega}_B)^4}
                                           ,  ~({\rm incorrect}),
\end{align}
which holds in the limit $|\nu|=|\omega- 2\tilde\omega_B|>v_F
q$. Importantly, the ``result'' is seemingly independent of a momentum
cut-off even though the intermediate steps do require an explicit
cut-off of the order $k_F$ when integrating over gauge momentum
$p_\perp$.

Carrying out the frequency integration first, similar to the
calculations in Maslov {\em et al.} \cite{maslov2018}, produces quite a different answer 
\begin{align}
 {\rm Re}\, \chi^1(\bm{q},\omega)  & \simeq \frac{k_F^2 q^2 (\omega + 2\tilde{\omega}_B)}{16 \pi^2 \chi (2\tilde{\omega}_B -\omega)^3} 
 \ln\left(\frac{\Lambda \chi}{\gamma}\right), ~({\rm correct}).
 \label{eq:o21}
 \end{align}
Here $\Lambda$ is a high-momentum cut-off. Observe that $ \Re \chi^1$ diverges in the $\gamma\to 0$ limit, which represents the static limit of gauge fluctuations, see \eqref{eq:12}.

The technicalities of integrations leading to Eq.~\eqref{eq:o21} are tedious and
require a number of simplifications performed at the proper stages in
the calculation.  These steps are carried out in the Mathematica
Notebook ``realchi1.nb" which is included in the Supplementary Materials, see the following link https://journals.aps.org/prb/supplemental/10.1103/PhysRevB.101.020401 . 
There we find that the most divergent part of the ${\rm Re} \chi^1(\bm{q},\omega)$ is given by 
\begin{align}
{\rm Re} \chi^1(\bm{q},\omega) = \frac{\chi^3 k_F^2 q^2 (\omega + 2\tilde{\omega}_B)}{16 \pi^3 (2\tilde{\omega}_B - \omega)^3} \int_0^{2\pi} d\phi \int_0^\Lambda dp \frac{p^7 \cos^2\phi}{(\gamma^2 (p - 2 k_F \sin\phi)^2 + \chi^2 p^4)^2}.
\label{eq:o22}
\end{align}
We rescale by pulling out $\chi$ from the denominator, introducing
$\alpha = \gamma/(\chi k_F)$ and letting $p = k_F y$. Then
\begin{align}
{\rm Re} \chi^1(\bm{q},\omega) = \frac{k_F^2 q^2 (\omega + 2\tilde{\omega}_B)}{16 \pi^3 \chi (2\tilde{\omega}_B - \omega)^3} \int_0^{2\pi} d\phi  \cos^2\phi \int_0^{\Lambda_F} dy \frac{y^7}{(\alpha^2 (y - 2 \sin\phi)^2 + y^4)^2},
\label{eq:o23}
\end{align}
where $\Lambda_F = \Lambda/k_F$ is the dimensionless upper cut-off of the momentum integration. We observe that for large $\Lambda_F \gg 1$ integration variable $y \sim \Lambda_F$ is large as well, and therefore the $y$-integral is approximated as $\int_0^{\Lambda_F} dy y^3/(\alpha^2 + y^2)^2 \approx \ln(\Lambda_F/\alpha)$.
This gives us the quoted result, Eq.\eqref{eq:o21}. 

In addition to the leading $q^2$ term in \eqref{eq:o21}, our calculation
also produces a finite zero momentum independent contribution, which
represents a correction $\delta M$ to the magnetization $M$
\begin{align}
\chi^1(0,\omega) =  \frac{2\delta M}{\omega - 2\omega_B} = \frac{1}{\omega - 2\omega_B} \frac{\omega_B m^2 \gamma}{2\pi^4 \chi^2 k_F} I(\gamma/\chi k_F).
\label{eq:o24}
\end{align}
Note that the prefactor is a simple pole, which reflects the Larmor theorem.
Here $I(\alpha)$ is a momentum integral, which we calculate numerically. It parametrically depends on its argument $\alpha = \gamma/\chi k_F$. For example, for $\alpha=1$ we obtain $I=0.76$.
Note that \eqref{eq:o24} represents a quantum correction to $M$ (and
hence the uniform susceptibility itself since $M$ is proportional to the
appleid field), and vanishes in the limit $\gamma \to 0$. Interestingly, this correction comes from the real part of $ \chi^{1}_{12,B}$, \eqref{eq:2}, and is entirely absent when one integrates over the momenta before integrating over the frequencies.

We conclude by noting that imaginary part of $\chi^1$, \eqref{eq:88},
is not sensitive to the order of integrations discussed here.
Carrying out calculations following \cite{Kim1994} and doing the
momentum integration before the frequency one precisely reproduces
Eq.~\eqref{eq:88} (we do not show those redundant calculations here).

\section{RPA form}
\label{sec:rpa}

In the RPA approximation 
\begin{align}
X_\pm(\bm{q},i \omega_n) = \frac{\chi^0_\pm(\bm{q},i \omega_n) + \chi^1_\pm(\bm{q},i \omega_n)}{1 + u [\chi^0_\pm(\bm{q},i \omega_n) + \chi^1_\pm(\bm{q},i \omega_n)]}
\label{eq:o25}
\end{align}
so that after the continuation to real frequency $i\omega_n \to \omega + i0$ we obtain for the imaginary part of the susceptibility outside the non-interacting particle-hole continuum
\begin{align}
{\rm Im}[X_\pm(\bm{q},\omega)] = \frac{{\rm Im}[\chi^1_\pm(\bm{q},\omega)]}{(1 + u~ {\rm Re}[\chi^0_\pm(\bm{q},\omega) + \chi^1_\pm(\bm{q},\omega)])^2+(u~ {\rm Im}[\chi^1_\pm(\bm{q},\omega)])^2}.
\label{eq:o26}
\end{align}
Damping of the collective spin mode is determined by \eqref{eq:88}, since ${\rm Im}[\chi^0_\pm(\bm{q},\omega)]=0$ in this frequency range, while its dispersion is given by 
\begin{align}
1 + u~ {\rm Re}[\chi^0_\pm(\bm{q},\omega) + \chi^1_\pm(\bm{q},\omega)] = 0.
\label{eq:o27}
\end{align}
Considering \eqref{eq:o21} as a small correction to the dispersion of the collective mode, which is justified by the $1/N$ expansion assumption of the gauge theory \cite{Kim1994}, we 
parameterize the dispersion as $\omega_q = 2\omega_B - \zeta q^2$ and solve \eqref{eq:o27} for $\zeta = 1/(2 m^*)$ by iterations. We find that the effective mass $m^*$ increases,
\begin{align}
\frac{1}{m^*} = \frac{v_F^2}{2u M} - \frac{u (2 \omega_B + uM)  \ln\big(\frac{\Lambda \chi}{\gamma}\big)}{4\pi^2 \chi (2 u M)^3},
\label{eq:o28}
\end{align}
and the dispersion of the collective mode flattens, correspondingly. But it retains its $q^2$ character.

\section{ESR in the U(1) Fermi surface spin liquid}
\label{sec:ESR-app}

\subsection{Mori-Kawasaki formula}
In the following we use Mori-Kawasaki (MK) formula, as derived in the
Appendix of \cite{Oshikawa2002}.  A calculation with Matsubara Green's
functions is also possible, and was in fact carried out -- the end
structure of result is the same in the two approaches.  This Matsubara
calculation is instructive for understanding which diagrams contribute
to the result.  It shows that the ESR response is
determined, in the leading $\alpha^2$ order, only by the diagrams with
self-energy insertions inside the spinon bubble.  The vertex diagram
is not present in that order. However, the calculation at finite
temperature $T$ requires an analytical continuation to real frequency
which is not an easy task.  The MK approach is formulated directly in
terms of retarded Green's functions which is more convenient.

Let us parameterize $H = H_0 + H' + H_Z$, where the first term is SU(2) invariant, the second contains symmetry-breaking perturbations and the third
describes interaction with static external field $H_Z = - h S^z_{\rm
  tot}$.    The {\em total} spin of the system $S^+$ obeys the
equations of motion
\begin{equation}
\frac{d S^+}{d t} = -i h S^+ + i {\cal R}, \frac{d S^-}{d t} = i h S^- - i {\cal R}^+,
\label{ap:5}
\end{equation}
where ${\cal R} = [H', S^+]$. 

Then for the retarded Green's function $G^R_{S^+ S^-}(\omega) = - i \int_0^\infty dt e^{i \omega t} \langle [S^+(t), S^-(0)]\rangle$ one obtains,
using the identity $e^{i\omega t} = \frac{1}{i\omega} \frac{d}{d t} e^{i\omega t}$ and integration by parts,
\begin{equation}
G^R_{S^+ S^-}(\omega) = \frac{2 \langle S^z\rangle}{\omega} + \frac{h}{\omega} G^R_{S^+ S^-}(\omega) - \frac{1}{\omega} G^R_{{\cal R} S^-}(\omega).
\label{ap:6}
\end{equation}
Repeating these steps for $G^R_{{\cal R} S^-}(\omega)$ one finds
\begin{equation}
G^R_{S^+ S^-}(\omega) = \frac{2 \langle S^z\rangle}{\omega - h +i0} - \frac{\langle[{\cal R},S^-]\rangle}{(\omega - h +i0)^2} + 
\frac{G^R_{{\cal R}{\cal R}^+}(\omega)}{(\omega - h +i0)^2} .
\label{ap:7}
\end{equation}
Assuming next that $G^R_{S^+ S^-}(\omega) = \frac{2 \langle
  S^z\rangle}{(\omega - h - \Sigma(\omega))}$ and expanding to second
order in the
perturbation ${\cal R} \sim \alpha$, we obtain explicit the expression for the self-energy $\Sigma(\omega)$
\begin{equation}
\Sigma(\omega) = \Big(\langle[{\cal R},S^-]\rangle - G^R_{{\cal R}{\cal R}^+}(\omega)\Big)/(2 \langle S^z\rangle)
\label{ap:8}
\end{equation}
The real part of this determines the shift of the resonance frequency
from the ideal $\omega = h = 2\omega_B$ value, while the imaginary
part determines all-important linewidth, resulting in
Eq.~\eqref{eq:55} of the main text.  Note that the first term in
Eq.~\eqref{ap:8} proportional to the commutator is frequency
independent and hence does not contribute to the imaginary part,
because the latter must be odd in frequency (as shown by the spectral
representation).

Direct calculation gives the result shown in the main text
\begin{equation}
{\cal R} = - i \frac{\alpha_{\rm so}}{2} \sum_{p,q} \psi^\dagger_{p+q}
\sigma^z \psi^{\vphantom\dagger}_{p}\, A^-(q).
\label{ap:10}
\end{equation}
This is a composite operator consisting of a convolution in momentum
space of $S^z(q) = \sum_p \psi^\dagger_{p+q} \frac{\sigma^z}{2} \psi^{\vphantom\dagger}_{p}$ and the gauge field $A^-(q)$,
${\cal R} = - i \alpha_{\rm so} \sum_q S^z(q) A^-(q)$.

\subsection{Convolution formula}
Since ${\cal R} \sim \alpha_{\rm so}$, the retarded Green's function
of the composite operator ${\cal R}$ can be approximated by
$\alpha_{so}^2$ multiplied by the simple convolution
of independent Green's functions of $S^z_q$ and $A^{-}_q$. This well known result is worked out below as follows.

Let $\mathcal{O} = A B$, where $A$ and $B$ stand for two arbitrary
independent (commuting) operators, $[A,B]=0$. 
Then
\begin{equation}
  \label{eq:57}
  G^R_{\mathcal{O}}(t) = - i\Theta(t) \langle [\mathcal{O}_t, \mathcal{O}_0^\dagger]\rangle = \Theta(t)( G^>_{\mathcal{O}}(t) - G^<_{\mathcal{O}}(t)),
\end{equation}
where the greater and lesser
Green's functions are defined by 
\begin{equation}
G^>_{\mathcal{O}}(t) = - i \langle \mathcal{O}_t
\mathcal{O}_0^\dagger\rangle, \qquad G^<_\mathcal{O}(t) =  i \langle \mathcal{O}_0^\dagger \mathcal{O}_t\rangle .
\label{ap:11}
\end{equation}
Note that here the lower index of the operators denotes time dependence, $\mathcal{O}_t = \mathcal{O}(t)$.

Using this representation and commutativity of $A$ and $B$ it is easy to show that 
\begin{equation}
G^R_{\mathcal{O}}(t) = i \Theta(t) \Big( G^>_A(t) G^>_B(t) - G^<_A(t) G^<_B(t) \Big).
\label{ap:12}
\end{equation}

Next, the spectral decomposition and $A_t = e^{i H t} A_0 e^{-i H t}$ lead to 
\begin{equation}
G^>_A(t) = \frac{-i}{Z} \sum_{n,m} |\langle m|A|n\rangle|^2 e^{-\beta E_m} e^{-it(E_n - E_m)} = \int d\epsilon e^{-i\epsilon t} 
\Big\{\frac{-i}{Z} \sum_{n,m} |\langle m|A|n\rangle|^2 e^{-\beta E_m} \delta(\epsilon - (E_n - E_m)) \Big\}
\label{ap:13}
\end{equation}
and
\begin{equation}
G^<_A(t) = \frac{-i}{Z} \sum_{n,m} |\langle m|A|n\rangle|^2 e^{-\beta E_n} e^{-it(E_n - E_m)} = \int d\epsilon e^{-i\epsilon t} e^{-\beta \epsilon}
\Big\{\frac{-i}{Z} \sum_{n,m} |\langle m|A|n\rangle|^2 e^{-\beta E_m} \delta(\epsilon - (E_n - E_m)) \Big\}.
\label{ap:14}
\end{equation}

As a result, spectral density (which determines $G^R_A(\omega)$) can be written as 
\begin{equation}
\rho_A(\epsilon) = \frac{1}{Z} \sum_{n,m} |\langle m|A|n\rangle|^2 (e^{-\beta E_m} - e^{-\beta E_n}) \delta(\epsilon - (E_n - E_m)).
\label{ap:15}
\end{equation}
It satisfies ${\rm sign}[\rho(\epsilon)] = {\rm sign}[\epsilon]$.

It is easy now to obtain
\begin{equation}
G^>_A(t) = (-i) \int d\epsilon e^{-i\epsilon t} \rho_A(\epsilon) [1 +
n(\epsilon)], \qquad G^<_A(t) = (-i) \int d\epsilon e^{-i\epsilon t} \rho_A(\epsilon) n(\epsilon).
\label{ap:16}
\end{equation}

This allows us to write \eqref{ap:12} as 
\begin{equation}
G^R_{\mathcal{O}}(t) = -i \Theta(t) \int d\epsilon_1 d\epsilon_2 e^{-i(\epsilon_1 + \epsilon_2)t} \rho_A(\epsilon_1) \rho_B(\epsilon_2) 
\{(1+n(\epsilon_1)) (1+n(\epsilon_2)) - n(\epsilon_1) n(\epsilon_2) \}.
\label{ap:17}
\end{equation}

Using $G^R(\omega) = \int_0^\infty dt e^{i(\omega + i0)t} G^R(t)$ we obtain
\begin{equation}
G^R_{\mathcal{O}}(\omega) =  \int d\epsilon_1 d\epsilon_2 \{1 + n(\epsilon_1) + n(\epsilon_2)\} 
\frac{\rho_A(\epsilon_1) \rho_B(\epsilon_2)}{\omega - \epsilon_1 -\epsilon_2 + i0},
\label{ap:18}
\end{equation}
so that
\begin{equation}
{\rm Im}G^R_{\mathcal{O}} (\omega) = -\pi \int d\epsilon_1 d\epsilon_2 \{1 + n(\epsilon_1) + n(\epsilon_2)\}  \rho_A(\epsilon_1) \rho_B(\epsilon_2)
\delta(\omega  - \epsilon_1 -\epsilon_2).
\label{ap:19}
\end{equation}

We can express this back in terms of the Green's function, because for
the individual retarded Green's function one has
\begin{equation}
G^R_A(\omega) = \int d\epsilon \frac{\rho_A(\epsilon)}{\omega - \epsilon + i0},
\label{ap:20}
\end{equation}
while the corresponding Matsubara Green's function $G^M_A(\tau) = - \langle T_\tau A_\tau A_0^\dagger\rangle$ is written as 
\begin{equation}
G^M_A(\omega_n) = \int d\epsilon \frac{\rho_A(\epsilon)}{i\omega_n - \epsilon}
\label{ap:21}
\end{equation}

Therefore 
\begin{equation}
{\rm Im}G^R_{\mathcal{O}} (\omega) = -\frac{1}{\pi} \int d\epsilon_1 d\epsilon_2 \{1 + n(\epsilon_1) + n(\epsilon_2)\}  {\rm Im}G^R_A(\epsilon_1) 
{\rm Im}G^R_B(\epsilon_2) \delta(\omega  - \epsilon_1 -\epsilon_2) .
\label{ap:22}
\end{equation}

\subsection{Calculation}

\subsubsection{Spectral densities}

And now we consider the relevant case here with $\mathcal{O} = AB$ and
 $A = S^z_q$ and $B = A^{-}_q$.   Note that we pulled out the factor
 of $\alpha_{\rm so}$ so $\mathcal{R} = \alpha_{\rm so} \mathcal{O}$.
 Ref.\onlinecite{Kim1994} tells us that 
$G^M_B(\nu) = \frac{q}{\gamma |\nu| + \chi q^3}$ so that $G^R_B(\epsilon) = \frac{q}{-i\gamma \epsilon + \chi q^3}$. Here 
$|\nu| \to -i \epsilon + 0$ is used for $\nu > 0$. Hence 
\begin{equation}
{\rm Im}G^R_B(\epsilon_2) = \frac{\gamma q \epsilon_2}{\gamma^2 \epsilon_2^2 + \chi^2 q^6} .
\label{ap:23}
\end{equation}

$G^M_A$ is the polarization bubble of two spinon lines,
\begin{equation}
G^M_A(\omega_n) = \int d\epsilon \int (d^2\bm{p}) \frac{f(\xi_p) - f(\xi_{p+q})}{i\omega_n - \epsilon} \delta(\epsilon + \xi_p - \xi_{p+q})
\label{ap:24}
\end{equation}
Standard angular integration gives
\begin{equation}
{\rm Im}G^R_A(\epsilon_1) = \frac{m}{2\pi} \frac{\epsilon_1}{\sqrt{v^2 q^2 - \epsilon_1^2}} \Theta(vq-|\epsilon_1|)
\label{ap:25}
\end{equation}

Thus we need to evaluate 
\begin{eqnarray}
{\rm Im}G^R_{\mathcal{O}} (\omega) &=& -\frac{1}{\pi} \int d\epsilon \int (d^2 \bm{q}) \{1 + n(\epsilon) + n(\omega-\epsilon)\}  
\frac{m}{2\pi} \frac{\epsilon}{\sqrt{v^2 q^2 - \epsilon^2}} \Theta(vq-|\epsilon|) 
\frac{\gamma q (\omega-\epsilon)}{\gamma^2 (\omega-\epsilon)^2 + \chi^2 q^6} \nonumber\\
&&=-\frac{m \gamma}{4\pi^3 v^3}  \int d\epsilon \{1 + n(\epsilon) + n(\omega-\epsilon)\} \int_{|\epsilon|}^\infty \frac{d q q^2}{\sqrt{q^2 - \epsilon^2}}
\frac{\epsilon(\omega-\epsilon)}{\gamma^2 (\omega-\epsilon)^2 + \frac{\chi^2 q^6}{v^6}}
\label{ap:26}
\end{eqnarray}

\subsubsection{$T=0$}
At zero temperature $1 + n(\epsilon) + n(\omega -\epsilon) \to \Theta(\epsilon) - \Theta(\epsilon-\omega) = \Theta(\epsilon)\Theta(\omega-\epsilon)$
and \eqref{ap:26} simplifies to
\begin{eqnarray}
{\rm Im}G^R_{\mathcal{O}} (\omega) &=& -\frac{m \gamma}{4\pi^3 v^3} \int_0^\omega d\epsilon   
\int_{|\epsilon|}^\infty \frac{d q q^2}{\sqrt{q^2 - \epsilon^2}}
\frac{\epsilon(\omega-\epsilon)}{\gamma^2 (\omega-\epsilon)^2 + \frac{\chi^2 q^6}{v^6}} = \nonumber\\
&&= -\frac{m}{4\pi^3 v^3 \gamma} \int_0^\omega d\epsilon \frac{\epsilon^3}{\omega - \epsilon} 
\int_1^\infty \frac{dx x^2}{\sqrt{x^2 -1}} \frac{1}{1+ \frac{\chi^2}{v^6} \big(\frac{\epsilon^3}{\omega - \epsilon}\big)^2 x^6}.
\label{ap:27}
\end{eqnarray}
The integration over $x$ contains no singularities and, for $\omega$
(hence also $\epsilon$) small, it is governed by large $x$, so we can
approximate $\sqrt{x^2 -1} \to x$ which allows one to evaluate it as 
\begin{equation}
\int \frac{dx x}{1+ \lambda x^6} \sim \lambda^{-1/3} = \frac{v^2}{\chi^{2/3}} \Big(\frac{\omega - \epsilon}{\epsilon^3}\Big)^{2/3} .
\label{ap:28}
\end{equation} 
The $\epsilon$-integration simplifies immediately as well
\begin{equation}
\int_0^\omega d\epsilon \Big(\frac{\epsilon^3}{\omega - \epsilon}\Big)^{1/3} = \omega^{5/3} \int_0^1 \frac{du u}{(1-u)^{1/3}} .
\label{ap:29}
\end{equation}
So we find ${\rm Im}G^R_{\mathcal{O}} (\omega) \sim - (m/(v \gamma
\chi^{2/3})) \omega^{5/3}$. Therefore, from Eq.~\eqref{eq:55} of the
main text, 
$\eta \sim \omega^{5/3}/M \sim \omega_B^{2/3}$ since $\langle
S^z\rangle = \chi_{\rm uniform} \omega_B$ and near the resonance $\omega = 2\omega_B$.

If no magnetic field is present, absorption is still possible because  spin-orbit interaction breaks spin-rotational
invariance.  In that case, at $T=0$, one has $\eta(\omega) \sim
\omega^{2/3}$. Such fractional dependence is quite familiar in 
U(1) gauge theory \cite{Kim1994}. 

\subsubsection{$T> 0$}
This case is more complicated. Eq.\eqref{ap:29} suggests that the
$\epsilon$-integration is dominated by the region where $\epsilon
\approx \omega$.  Let $\epsilon = \omega + \nu$ and approximate 
$n(-\nu) \approx -T/\nu$, then the leading part of \eqref{ap:26}
reduces to
\begin{eqnarray}
{\rm Im}G^R_{\mathcal{O}} (\omega) &&= -\frac{m \gamma}{4\pi^3 v^3} \int d\nu \frac{-T}{\nu} \int_{|\omega|}^\infty \frac{d q q^2}{\sqrt{q^2 - \omega^2}}
\frac{\omega (-\nu)}{\gamma^2 \nu^2 + \frac{\chi^2 q^6}{v^6}} = -\frac{m T \omega}{4\pi^2 \chi} \int_{|\omega|}^\infty \frac{d q}{q\sqrt{q^2 - \omega^2}}
\nonumber\\
&& =-\frac{mT {\rm sign}(\omega)}{4\pi^2 \chi} \int_1^\infty \frac{dx}{x\sqrt{x^2-1}}.
\label{ap:30}
\end{eqnarray}
Here we approximated $\epsilon \to \omega$ in all places which do not
cause any divergence. The $\omega$-dependence has dropped out!
The $x$-integral is actually $\pi/2$.

This important observation suggests the following manipulation of \eqref{ap:26}: add and subtract $T/(\omega - \epsilon)$ to separate the 
linear in $T$ piece obtained above from the expected $T^{5/3} f(\omega/T)$ scaling part. So,
$\{1 + n(\epsilon) + n(\omega-\epsilon)\} = [1 + n(\epsilon) + n(\omega-\epsilon) - \frac{T}{\omega - \epsilon}] + \frac{T}{\omega - \epsilon}$.
The last term on the right gives \eqref{ap:30}. 

We now look at the first one, coming from the expression inside square brackets. Denote this contribution to ${\rm Im}G^R_{\mathcal{O}} (\omega)$ as
$-C$. Then
\begin{equation}
C = \frac{m \gamma}{4\pi^3 v^3}  \int d\epsilon [1 + n(\epsilon) + n(\omega-\epsilon) - \frac{T}{\omega - \epsilon}] \int_{|\epsilon|}^\infty \frac{d q q^2}{\sqrt{q^2 - \epsilon^2}}
\frac{\epsilon(\omega-\epsilon)}{\gamma^2 (\omega-\epsilon)^2 + \frac{\chi^2 q^6}{v^6}}
\label{ap:31}
\end{equation}
Now we change to scaling variables: $\omega = T x, \epsilon = T y, q = T^{1/3} p$. Then $\sqrt{q^2 - \epsilon^2} = \sqrt{T^{2/3} p^2 - T^2 y^2} \to T^{1/3} p$.
The $p$-integral reduces to 
\begin{equation}
\frac{T^{2/3}}{\gamma^2} \int_0^\infty dp p \frac{y(x-y)}{(x-y)^2 + \frac{\chi^2}{\gamma^2 v^6} p^6} = \frac{T^{2/3}}{\gamma^2}
\big(\frac{\gamma v^3}{\chi}\big)^{2/3} y (x-y) \frac{|x-y|^{2/3}}{(x-y)^2} \int_0^\infty \frac{dz z}{1+z^6}.
\label{ap:32}
\end{equation}
The $z$-integral is just $\pi/(3\sqrt{3})$. This turns \eqref{ap:31} into
\begin{equation}
C = \frac{m T^{5/3}}{12 \sqrt{3} \pi^2 v \gamma^{1/3} \chi^{2/3}}  \int_{-\infty}^\infty dy [\frac{e^y}{e^y-1} + \frac{1}{e^{x-y}-1} - \frac{1}{x-y}] 
\frac{y(x-y)}{|x-y|^{4/3}}
\label{ap:33}
\end{equation}
We split the $y$-integral into two, according to the sign of $|x-y|$, $C = C_a + C_b$. Hence
\begin{equation}
C_a = c_0 \int_{-\infty}^x dy [...] \frac{y}{(x-y)^{1/3}}, C_b = (- c_0) \int_x^{\infty} dy [...] \frac{y}{(y-x)^{1/3}}, 
\label{ap:34}
\end{equation}
where $c_0 = \tilde{c}_0 T^{5/3}$ is an overall constant in
\eqref{ap:33}, i.e.
\begin{equation}
  \label{eq:58}
  \tilde{c}_0 = \frac{m}{12 \sqrt{3} \pi^2 v \gamma^{1/3} \chi^{2/3}} .
\end{equation}
In the $C_a$, we change $y = 2x - y'$ so that $x-y = y' - x$, and find
\begin{equation}
C_a = c_0 \int_x^{\infty} dy [\frac{e^{2x-y}}{e^{2x-y}-1} + \frac{1}{e^{y-x}-1} + \frac{1}{x-y}] \frac{2x-y}{(y-x)^{1/3}}
\label{ap:35}
\end{equation}
where we instantly renamed $y' \to y$ again. We now add $C_b$ and simplify to find
\begin{equation}
C = c_0 \int_x^\infty \frac{dy}{(y-x)^{1/3}} \Big( \frac{2x-y}{1-e^{y-2x}} - \frac{y}{e^y-1} + 2x[\frac{1}{e^{y-x}-1} - \frac{1}{y-x}]\Big) .
\label{ap:36}
\end{equation}
This expression is free from divergences. Finally, setting $y = z+ x$ we obtain
\begin{equation}
{\rm Im}G^R_{\mathcal{O}} (\omega) = - \tilde{c}_0 T^{5/3} f(x) = - \tilde{c}_0 T^{5/3} \int_0^\infty \frac{dz}{z^{1/3}} 
\Big( \frac{x-z}{1-e^{z-x}} - \frac{z+x}{e^{z+x}-1} + 2x[\frac{1}{e^z-1} - \frac{1}{z}]\Big) .
\label{ap:37}
\end{equation}
This represents the desired scaling function, with $x=\omega/T$. The plot of the scaling function $f(x)$ is shown in Figure \ref{fig:eta}.
Its limits are as follows:
\begin{equation}
f(x) \to -4.37746 x ~{\rm for}~ x\ll 1; f(x) \to 0.75 x^{5/3} ~{\rm for}~ x\gg 1.
\label{ap:38}
\end{equation}
The small-$x$ limit works for $x \leq 5$ while the large-$x$ limit turns on only for really large $x \sim 100$. 
Altogether we have
\begin{equation}
\eta = \frac{\alpha_{\rm so}^2}{2\chi_{\rm uniform} \omega_B} \Big(\frac{m T}{8\pi \chi} + \tilde{c}_0 T^{5/3} f(\frac{2\omega_B}{T})\Big) = \frac{\alpha_{\rm so}^2}{2\chi_{\rm uniform}} \Big(\frac{m}{8\pi \chi} \frac{T}{\omega_B} + \tilde{c}_0 T^{2/3} \tilde{f}(\frac{2\omega_B}{T})\Big) ,
\label{ap:39}
\end{equation}
where we used the resonance condition $\omega = 2\omega_B$. Also here $\tilde{f}(x) = f(x)/x$.
The interesting $T^{2/3}$ behavior of $\eta$ is clearly subleading.

\end{document}